%% file: article.tex
\documentclass[preprint]{siamart0516}

\input{shared}

\hypersetup{
  pdftitle={\TheTitle},
  pdfauthor={\TheAuthors}
}
\begin{document}
\parindent=0pt
\maketitle

\begin{abstract}
We present a new algorithm for the automatic one-shot generation of scattered node sets on irregular 2D and 3D domains using Poisson disk sampling coupled to novel parameter-free, high-order parametric Spherical Radial Basis Function (SBF)-based geometric modeling of irregular domain boundaries. Our algorithm also automatically modifies the scattered node sets \emph{locally} for time-varying embedded boundaries in the domain interior. We derive complexity estimates for our node generator in 2D and 3D that establish its scalability, and verify these estimates with timing experiments. We explore the influence of Poisson disk sampling parameters on both quasi-uniformity in the node sets and errors in an RBF-FD discretization of the heat equation. In all cases, our framework requires only a small number of ``seed'' nodes on domain boundaries. The entire framework exhibits $O(N)$ complexity in both 2D and 3D. 
\end{abstract}
\begin{keywords}
Node generation, Geometric Modeling. Radial Basis Functions, Poisson disk sampling, Embedded Boundaries.
\end{keywords}
% REQUIRED
\begin{AMS}
  68Q25, 68U05
\end{AMS}

\input{Introduction}
\input{Overview}
\input{NodeGeneration}

\input{Discussion}
\section*{Acknowledgments}
The authors were supported by NSF DMS-1521748. VS was also supported by DMS-1160432.

%\appendix
%\input{Appendix}
\bibliographystyle{siamplain}
\bibliography{article}

\end{document}

%% file: shared.tex
\usepackage{lineno}
\usepackage{amssymb,amsxtra,amsmath}
\usepackage{latexsym}
\usepackage{ifthen}
\usepackage{graphicx}
\usepackage{calc,fancyhdr}
\usepackage{indentfirst}
\usepackage{parskip,algorithm,algorithmicx}
\usepackage{algpseudocode}
\usepackage{color}
\usepackage{bigints}
\usepackage{subfig}
\usepackage{booktabs}
\usepackage{pdfpages}
\usepackage{Article}

\newcommand{\TheTitle}{Robust Node Generation for Meshfree Discretizations on Irregular Domains and Surfaces} 
\newcommand{\TheAuthors}{Varun Shankar, Robert. M. Kirby, and Aaron L. Fogelson}

\headers{Robust node generation for meshfree methods}{\TheAuthors}

\title{{\TheTitle}\thanks{Submitted to the editors 07/26/2017.}}

\author{
  Varun Shankar\thanks{Department of Mathematics, University of Utah, Salt Lake City, UT USA
    (\email{vshankar@math.utah.edu}, \url{http://www.math.utah.edu/\string~vshankar/}). Corresponding author.}
  \and  
  Robert M. Kirby\thanks{School of Computing and Scientific Computing and Imaging Institute, University of Utah, Salt Lake City, UT USA (\email{kirby@cs.utah.edu},
    \url{http://www.cs.utah.edu/\string~kirby/}).}
	\and
  Aaron L. Fogelson\thanks{Departments of Mathematics and Bioengineering, University of Utah, Salt Lake City, UT USA (\email{fogelson@math.utah.edu},
    \url{http://www.math.utah.edu/\string~fogelson/}).}		
}

%% file: Introduction.tex
\section{Introduction}
\label{sec:intro}

Many engineering applications require solving partial differential equations (PDEs) on both surfaces and their enclosed volumes. A typical pipeline is that data drive one's understanding of the geometry, and from that initial data a discretization of the PDE is built-- either via mesh-based methods (e.g. finite element methods (FEM)) or from the class of mesh-free methods (e.g. radial basis function (RBF) methods).  From the latter class, RBF-based Finite Difference (RBF-FD) methods have been used for over a decade to solve PDEs~\cite{Bayona2010,Davydov2011,CecilQian2004,Wright200699,Chandhini2007,FlyerWright:2007,FlyerWright:2009,FoL11,FlyerLehto2012,Piret2012, Piret2016,FuselierWright2013,SWFKJSC2014}, and recent work by Flyer and coworkers~\cite{FlyerPHS,FlyerNS,BarnettPHS} has helped overcome the traditional problems of RBF interpolants. Thus, for the remainder of this article, we will assume PDE discretizations based on RBF-FD methods.

One motivation for developing the methods described in this paper is our own work in simulating the blood clotting, which involves the formation of an aggregate of platelets. We treat each platelet as a discrete object (reconstructed from a set of discrete points) moving with the fluid in which it is immersed~\cite{Fogelson2008, SWFKAPNUM2013, SWFKIJNMF2015}. Proteins and other molecules that are secreted into the fluid by platelets mediate interplatelet interactions. The fluid concentrations of these molecules each satisfy an advection-diffusion-reaction equation in the fluid domain, with Robin-type boundary conditions satisfied on each platelet's surface. There may be several hundred platelets in the overall domain, with their numbers changing over time as new platelets enter and exit the domain; the fluid domain (that part of the overall domain external to all platelets) itself changes as the platelets' positions change. In order to apply RBF-FD to this problem, a rapid and \emph{local} node generation and modification algorithm is required. The methods presented in this paper may also be valuable in constructing geometry and node sets for solving PDEs from medical images; such images are used to define the geometry of a vascular bed for simulations of blood flow in that vascular bed~\cite{Steinman02,TF09}.  While current methods utilize meshing and an FEM-based solution framework, a rapid node generation and modification algorithm would allow direct solution of PDEs on the point-cloud data obtained from these images using RBF-FD. Another application which advances in cellular imaging should make possible soon is constructing intracellular domains, accounting for the presence of cellular organelles, and generating node sets within them for the solution of fluid-flow or advection-diffusion-reaction equations for intracellular processes~\cite{HHM14,RMS13}.

With the above applications in mind and from by previous experience with RBF-FD~\cite{ShankarJCP2017}, we attempted to generate node sets using existing tools such as Gmsh~\cite{GMSH} and Distmesh~\cite{Persson_distmeshpaper}, with the idea that we would generate surface and volumetric meshes, disregard the edge information, and use the remaining node set for building differential operators. However, numerical differential operators built on these node sets were unstable on irregular domains (e.g., approximations of second-order differential operators that had positive real eigenvalues). We then turned to the graphics community for inspiration on point-based sampling techniques, and tried Poisson disk sampling~\cite{Bridson07,Yuksel2015}. A naive implementation (without careful tuning) of Poisson disk sampling also resulted in point sets that were bunched at curved boundaries and/or that generated poorly-conditioned numerical operators.  

To overcome these challenges, we designed node generation algorithms (based on Poisson disk sampling) for irregular domains using only a small set of locations on domain boundaries/surfaces as a starting point. The algorithms presented in this paper attempt to satisfy four criteria. First, since RBF-FD is our chosen high-order discretization method, we require node sets on domain boundaries and the enclosing volume that approximately respect a user-specified node spacing $h$ on the surface and the volume; convergence estimates for RBF-FD methods are typically specified in terms of such measures of node spacing~\cite{DavydovSchaback2016,Fasshauer:2007}. Second, we wish to eliminate free or tuning parameters (other than $h$) for generating these node sets in order to provide mathematical modelers with robust and automatic tools for solving their model equations. Third, in order to rapidly generate node sets on time-varying domains, we desire \emph{one-shot} node generation algorithms with the ability to modify node sets \emph{locally}, in contrast to existing (iterative) repulsion-based approaches~\cite{fornbergflyerfast}. Fourth, in order to reduce time-to-discovery, we want our methods to be (provably) computationally efficient and scalable with regards to the number of points needed to discretize the problem, regardless of the order of the RBF-FD method used. \revone{Our focus in this work is not on generating a suitable parametrization for an arbitrary point cloud, which is a problem that has been tackled by others (\emph{e.g.},~\cite{ChoiHo2016}). We restrict ourselves to irregular domains with boundaries that are easily parameterizable, of genus-0, and at least homeomorphic to the sphere $\mathbb{S}^{d-1} \subset \mathbb{R}^d$. }

The remainder of our paper is structured as follows.  In Section \ref{sec:geom_models}, we present a robust approach to geometric modeling using Spherical RBF (SBF) interpolation, and confirm its high-order convergence rates with numerical experiments. In Section \ref{sec:node_gen}, we present a node generation pipeline that utilizes this new geometric modeling technique and Poisson disk sampling to generate scattered node sets on irregular domains and surfaces while satisfying our four design criteria. We also verify the quasi-uniformity of these node sets via histograms, and test their influence on stability and errors in high-order RBF-FD discretizations. In Section~\ref{sec:complex}, we derive complexity estimates that show the scalability of our proposed scheme, and verify these estimates with timing experiments.  We conclude in Section \ref{sec:summary} by discussing the trade-offs of our work and possible future directions.

\textbf{A note on symbols}: For readers' convenience, we have listed and defined the important symbols used in this article in Table \ref{tab:symbols}. This table is not exhaustive, and other symbols will be defined in the text where required.
\begin{table}[h!]
\footnotesize
\centering
\begin{tabular}{cc} \toprule
		Symbol & Meaning \\ \midrule
		$N$ & Total number of domain nodes\\
		$N_i$ & Number of interior nodes in the domain \\
		$N_b$ & Number of boundary nodes \\
		$N_{obb}$ & Number of nodes in the domain bounding box\\
		$N_d$ & Number of ``seed nodes'' and data sites for geometric model\\
		$q$ & Length/perimeter of the domain boundary in 2D\\
		$a$ & Area of the domain in 2D\\
		$a_{obb}$ & Area of the domain bounding box in 2D\\
		$s$ & Surface area of the domain boundary in 3D\\
	  $v$ & Volume of the domain in 3D\\
		$v_{obb}$ & Volume of the domain bounding box in 3D\\
		$h$ & Approximate node spacing in domain \\
		$h_i$ & Approximate node spacing in domain interior \\
		$h_b$ & Approximate node spacing on domain boundary \\		
		$N_p$ & Number of embedded inner boundaries\\
		$N_b^{\Gamma}$ & Number of nodes on each embedded boundary\\
		$N_d^{\Gamma}$ & Number of seed nodes on each embedded boundary\\
		$q^{\Gamma}$ & Length/perimeter of each embedded boundary in 2D\\
		$s^{\Gamma}$ & Surface area of each embedded boundary in 3D\\
		$h_{\Gamma}$ & the node spacing on each embedded boundary \\ \bottomrule		
\end{tabular}
\caption{Table of symbols.}
\label{tab:symbols}
\end{table}

%% file: Overview.tex
\section{Geometric Modeling with RBFs}
\label{sec:geom_models}

\revone{The foundation of our node generation algorithm for irregular domains is a method for reconstructing the domain boundary from a set of ``seed'' nodes obtained from an application. One approach to doing this reconstruction is to form a parametric geometric model of the domain boundary.} The authors have previously developed parametric geometric models based on \emph{global} spherical RBF (SBF) interpolants that have been used for modeling/reconstructing irregular surfaces in several applications~\cite{SWFKAPNUM2013,SWFKIJNMF2014,SWFKIJNMF2015,ShankarOlson2015,FuselierWright2013,FHNWW2013a}. \comment{However, these methods require the user to tune a shape parameter. We now present a modification of the geometric models for closed surfaces developed in~\cite{SWFKAPNUM2013} that eliminates this parameter. We assume the closed surface $\mathbb{M} \subset \mathbb{R}^d$ is homeomorphic to the unit sphere $\mathbb{S}^{d-1}$, where $d$ is the dimension of the embedding space. Let $\chid = \{\vX_k\}_{k=1}^{N_d}$ be a set of \emph{data sites} on $\mathbb{M}$. For the following discussion, we assume that $\Lambda = \{\vlambda_k\}_{k=1}^{N_d} = \{\lambda_k^1, \lambda_k^2, \hdots, \lambda_k^{d-1}\}_{k=1}^{N_d}$ is the set of parameter values for the data sites $\chid$.

\subsection{Modeling closed surfaces}
We focus on the problem of modeling a closed surface $\mathbb{M} \subset \mathbb{R}^d$. Let $\vxi(\vlambda) \in \mathbb{S}^{d-1}$. Furthermore, assume we are given $N_d$ points $\vX_1, \hdots, \vX_{N_d}$ on $\mathbb{M}$, where $\vX_k = \{X^1_k,\hdots,X^d_k\}$. To model the surface, we use an SBF interpolant generated from a polyharmonic spline (PHS) kernel.  Interpolating $X^j$ with this new PHS SBF interpolant, we have
\begin{align}
s^j(\vlambda)= \sum\limits_{k=1}^{N_d} c^j_k \phi\lf(\sqrt{2 (1 - \xi(\vlambda)\cdot\xi(\vlambda_k))}\rt),
\label{eq:sbf_poly_model}
\end{align}
where $\phi(r) = r^m$ on $\mathbb{S}^1$ ($m$ is odd), and $\phi(r) = r^m \log (r)$ on $\mathbb{S}^2$ ($m$ is even). For $\mathbb{S}^1$ (the unit circle), $\vlambda = \lambda^1 = \lambda$, allowing us to write \eqref{eq:sbf_poly_model} as:
\begin{align}
s^j(\lambda)= \sum\limits_{k=1}^{N_d} c^j_k \lf(2 - 2\cos\lf(\lambda - \lambda_k \rt)\rt)^\frac{m}{2}, 0<m\notin 2\mathbb{N}.
\label{eq:sbf_2d}
\end{align}
A similar (albeit more complicated) expression can be derived for an interpolant on $\mathbb{S}^2$ (the unit sphere). To find the coefficients, we solve the following linear systems for $j=1,\hdots,d$:
\begin{align}
\begin{bmatrix}
\lf(2 (1 - \xi(\vlambda_1)\cdot\xi(\vlambda_1))\rt)^\frac{m}{2} & \hdots & \lf(2 (1 - \xi(\vlambda_1)\cdot\xi(\vlambda_{N_d}))\rt)^\frac{m}{2} \\
\lf(2 (1 - \xi(\vlambda_2)\cdot\xi(\vlambda_1))\rt)^\frac{m}{2} & \hdots & \lf(2 (1 - \xi(\vlambda_2)\cdot\xi(\vlambda_{N_d}))\rt)^\frac{m}{2} \\
\vdots & \ddots & \vdots \\
\lf(2 (1 - \xi(\vlambda_{N_d})\cdot\xi(\vlambda_1))\rt)^\frac{m}{2} & \hdots & \lf(2 (1 - \xi(\vlambda_{N_d})\cdot\xi(\vlambda_{N_d}))\rt)^\frac{m}{2}
\end{bmatrix}
\begin{bmatrix}
c^j_1 \\
\vdots \\
c^j_{N_d}
\end{bmatrix} =
\begin{bmatrix}
X^j_1 \\
\vdots \\
X^j_{N_d}
\end{bmatrix}.
\label{eq:sbf_linsys}
\end{align}
In the case of standard PHS RBF interpolants, it is important to augment the RBFs with polynomials of degree $\lf\lfloor\frac{m-1}{2}\rt\rfloor$ (or higher) to regularize the end conditions~\cite{FornbergWrightRunge}. However, in the case of PHS SBFs, this is unnecessary due to  periodicity. \revtwo{More important is the question of the invertibility of \eqref{eq:sbf_linsys}}. It is well-known that the inclusion of polynomials is required to show unisolvency of the interpolant for conditionally positive-definite kernels such as the PHS kernel~\cite{Fasshauer:2007}. However, in our experiments, we found that we were able to solve the linear system in \eqref{eq:sbf_linsys} in all cases with simple Gaussian elimination even without the inclusion of trigonometric polynomials. 

It is important to note that unlike in~\cite{SWFKAPNUM2013} where infinitely-smooth SBFs with shape parameters were used, we use the piecewise-smooth PHS SBF with a fixed $m$, eliminating the need for tuning the shape parameter; we remark on the choice of $m$ in Section \ref{sec:geom_res}. We demonstrate in Section \ref{sec:geom_res} that this setting allows for spectral convergence rates when recovering smooth functions, despite the use of a piecewise-smooth SBF. The interpolation matrix in \eqref{eq:sbf_linsys} can be decomposed for a cost of $O(N_d^3)$, with subsequent coefficient calculations for all functions of the parametric nodes costing $O(N_d^2)$ operations.} We will discuss the computational complexity of this technique in the context of node generation in a later section.

\subsubsection{\revone{Parametrization and Interpolation nodes}}
\label{sec:node_sets_closed}

\revone{We now discuss the parametrization of the seed nodes on the domain boundary. This parametrization will be used within the SBF interpolant \eqref{eq:sbf_poly_model}. Clearly, since the SBF interpolant assumes parametrization on $\mathbb{S}^{d-1}$, we must parametrize the seed nodes on the same set. However, as mentioned in Section \ref{sec:intro}, this is in general a non-trivial task in 3D. For the sake of this article, we focus on simple parametrizations of the seed nodes on the domain boundary.}

When modeling a closed surface $\mathbb{M} \subset \mathbb{R}^2$, a natural parametric node set for interpolation is $\Lambda = \{\lambda_k\}_{k=1}^{N_d} \in [-\pi,\pi)$. For the purposes of this article, we restrict ourselves to equispaced samples. If adaptive sampling is required, these samples can be clustered appropriately.

For $\mathbb{M} \subset \mathbb{R}^3$, we use the parametric node set $\Lambda = \{\vlambda_k\}_{k=1}^{N_d} = \{\lambda_k,\theta_k\}_{k=1}^{N_d}$, where $-\pi \leq \lambda < \pi$ and $-\pi/2 \leq \theta <\pi/2$. However, obtaining the $(\lambda_k,\theta_k)$ pairs is not quite as simple as in the case of $\mathbb{M} \subset \mathbb{R}^2$. While equispaced points in $[-\pi,\pi)$ correspond to equispaced Cartesian samples for $\mathbb{S}^1$, this is not true for $\mathbb{S}^2$. \revone{As was done previously~\cite{SWFKAPNUM2013}, we use quasi-uniform Cartesian nodes on the sphere transformed into spherical coordinates in the rectangle $[-\pi,\pi)\times [-\pi/2, \pi/2)$ to obtain our parametric node sets}. There are many quasi-uniform node sets used for this purpose in the RBF literature, such as minimal energy (ME) or maximal determinant (MD) nodes~\cite{SWFKAPNUM2013}. However, these node sets are typically obtained by solving expensive optimization problems. Other node sets, such as icosahedral nodes, are only available for certain values of $N_d$. We select a node set that is easily computed on-the-fly in $O(N_d)$ operations for any value of $N_d$: the so-called \emph{generalized spiral points}, developed by Rakhmanov, Saff, and Zhu~\cite{SaffSpiral}. Interestingly, the algorithm presented in~\cite{SaffSpiral} was improved by Knud Thomsen in an online discussion group~\cite{SaffPage,ThomsenGGPage}. We opt to use Thomsen's generalized spiral algorithm, presented in simple form on~\cite{SaffPage}. \revone{Once the generalized spiral points are generated, we then transform them into spherical coordinates to obtain the parametric node set $\Lambda$.}

\subsubsection{Evaluation nodes}
\label{sec:node_sets_eval}
While a good choice of interpolation nodes controls geometric modeling error, the choice of evaluation nodes affects the errors in RBF-FD discretizations of PDEs on surfaces and domain boundaries. The goal is to find a set of parametric evaluation nodes in the rectangle $[-\pi,\pi)\times [-\pi/2, \pi/2)$ (or in $[-\pi,\pi]$ for curves)  that result in quasi-uniformly spaced Cartesian nodes on the surface. While this can be accomplished by solving optimization problems or rejection sampling in parametric space, we will adopt an approach that is a combination of parametric supersampling and \emph{Cartesian} thinning. This is explained in the context of our node generator in the next section.

\subsection{Convergence of geometric models}
\label{sec:geom_res}
\begin{figure}[h!]
\centering
\subfloat[Approximating a smooth function]
{
	\includegraphics[scale=0.4]{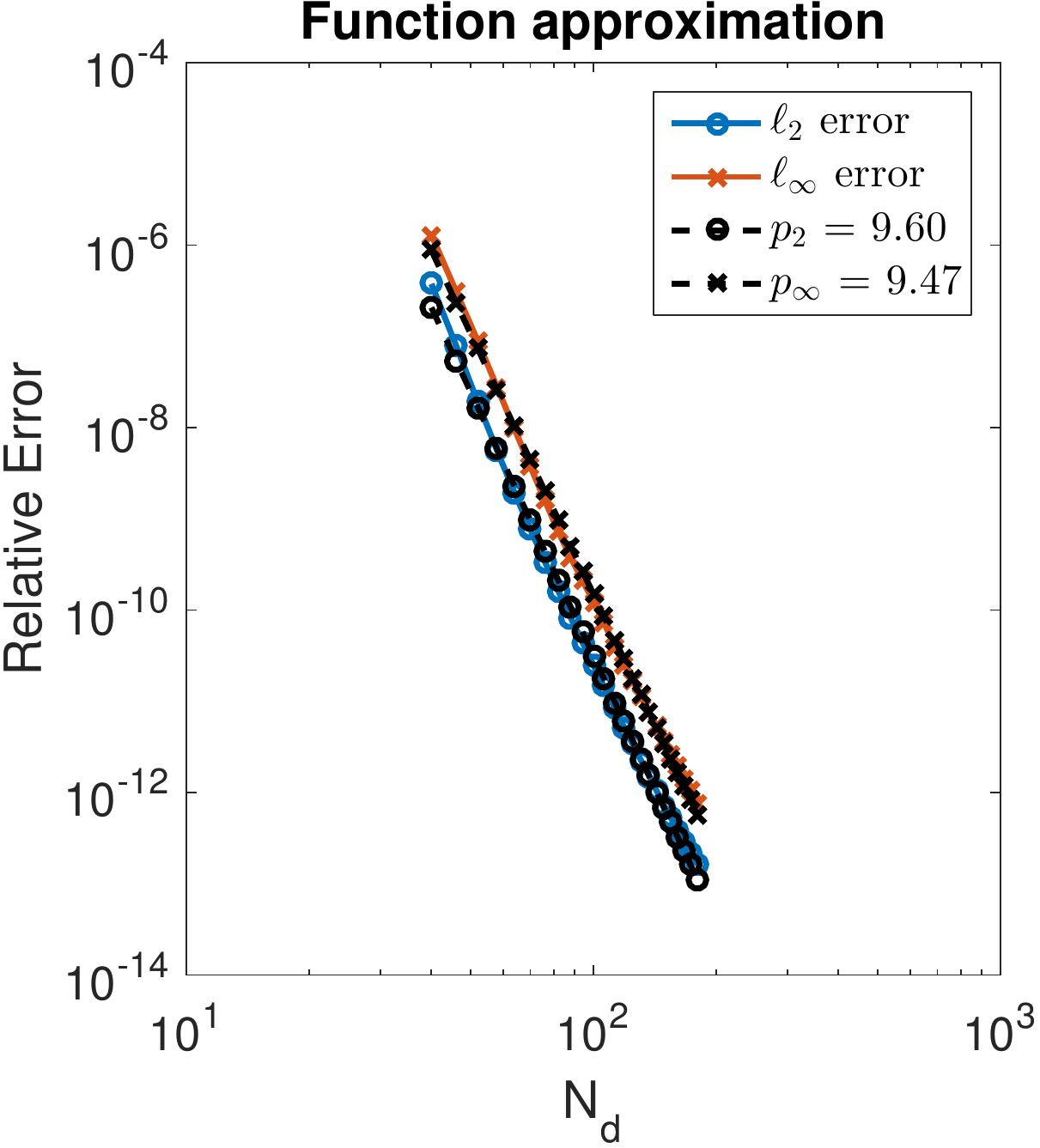} 	
	\label{fig:1a}
}
\subfloat[First derivative]
{
	\includegraphics[scale=0.4]{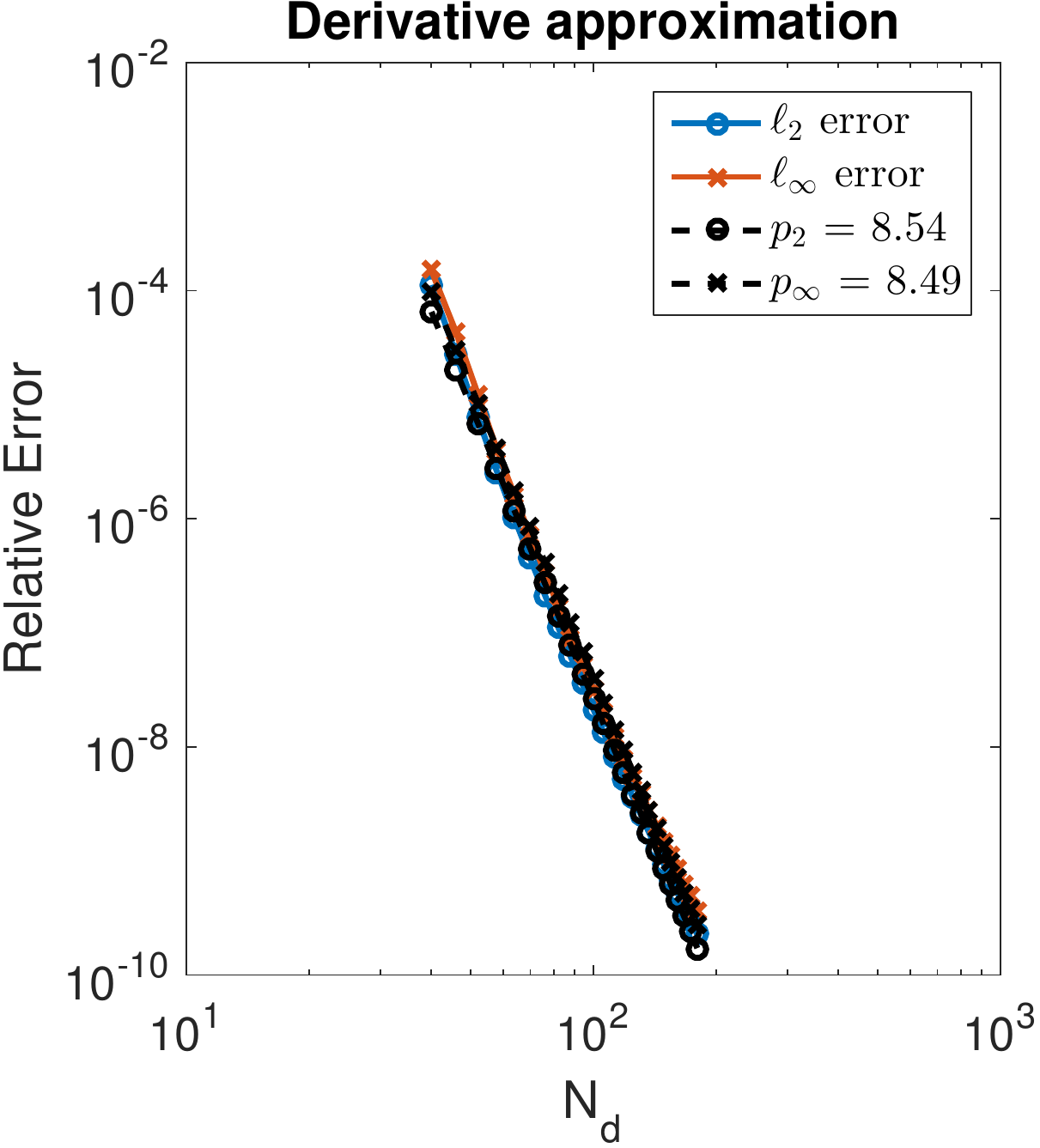}
	
	\label{fig:1b}
}

\subfloat[Approximating a $C^2$ function]
{
	\includegraphics[scale=0.4]{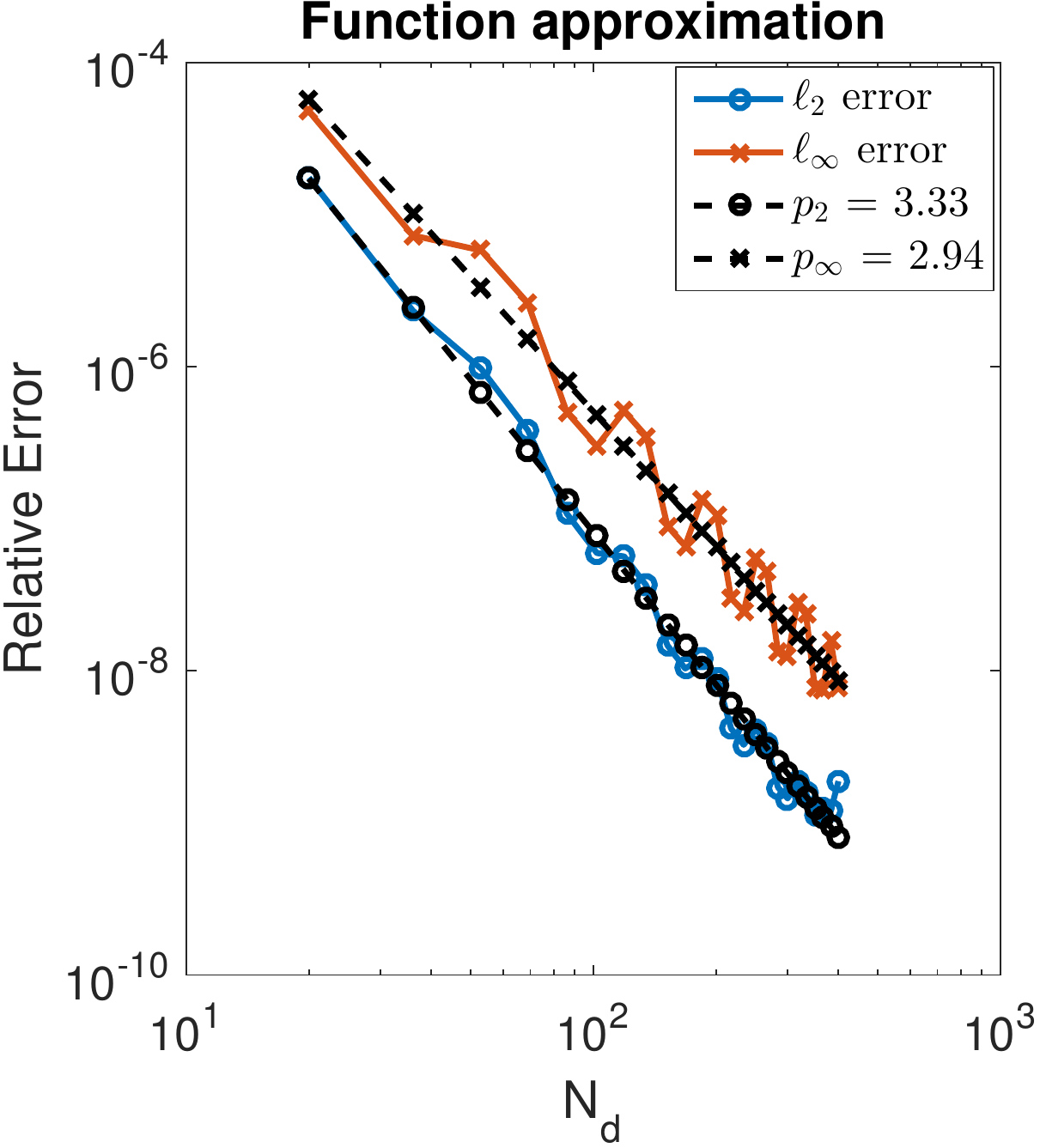} 	
	\label{fig:1c}
}
\subfloat[First derivative]
{
	\includegraphics[scale=0.4]{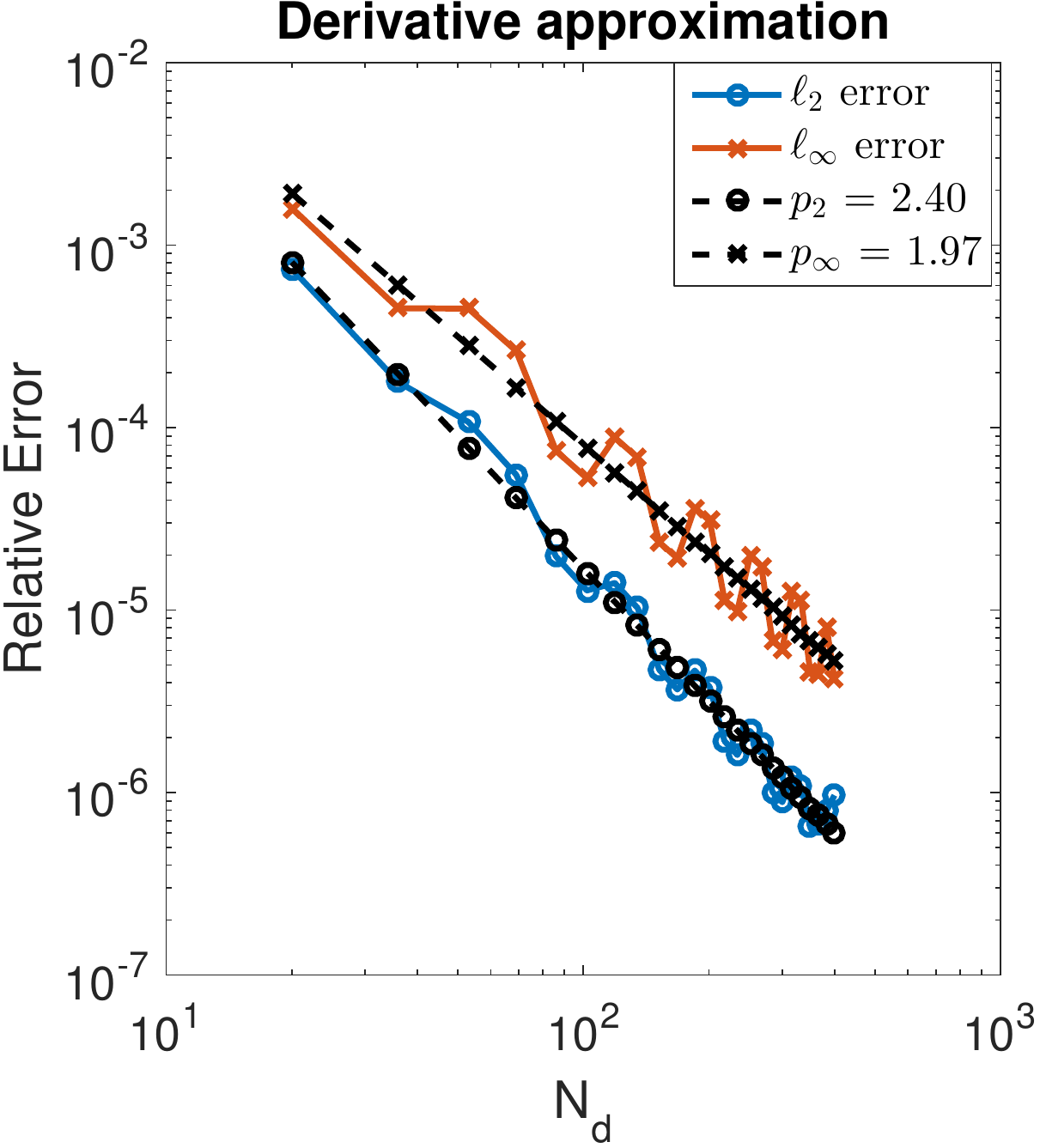} 	
	\label{fig:1d}
}
\caption{Errors in function and derivative approximation for closed 1D curves using $\phi(r) = r^7$. The dashed lines in the figures on the bottom row are lines of best fit with slopes shown in the legends.}
\label{fig:1Dresults}	
\end{figure}
In this section, \revone{we test the convergence of our new geometric model for closed surfaces with the goal of understanding its behavior on reconstructing both infinitely-smooth and finitely-smooth domain boundaries}. We use the boundaries presented in~\cite{SWFKAPNUM2013}, \revone{repeated here for reference. The functions corresponding to the 1D curves are given by:
{\small
\begin{align}
\textit{2D } C^{\infty} &: \lf[1 + A \rm{exp}\lf(\frac{-(1-\cos\lambda)^2}{\sigma_1} \rt) \rt] \vx_{ideal}, \\
\textit{2D } C^{2} &:\lf[1 + B \rm{exp}\lf(\frac{-(1-\cos^2\lambda)^{1.5}}{\sigma_2} \rt) \rt] \vx_{ideal}, \\
\vx_{ideal}(\lambda) &= \lf(x_c + a \cos\lambda, y_c + b\sin\lambda \rt),
\end{align}}
\noindent where $-\pi \leq \lambda \leq \pi$. For the boundary represented by the $C^{\infty}$ function, we use $x_c = y_c = 0.9$, $a=0.04$, $b=0.05$, $A=0.09$, and $\sigma_1 = 0.1$. For the boundary represented by the $C^2$ function, we use $x_c = y_c = 0.2$, $a=b=0.1$, $B=0.04$, and $\sigma_2 = 0.9$. The functions corresponding to the 2D surfaces are given by:
{\small
\begin{align}
\textit{3D } C^{\infty} &: \lf[1 + A \rm{exp}\lf(\frac{r_c^2}{\sigma_1} \rt) \rt] \vx_{ideal}, \\
\textit{3D } C^{3} &:\lf[1 + B \rm{exp}\lf(\frac{r_c^{1.5}}{\sigma_2} \rt) \rt] \vx_{ideal}, \\
\vx_{ideal}(\lambda) &= \lf(x_c + a \cos\lambda\cos\theta, y_c + b\sin\lambda\cos\theta, z_c + c\sin\theta \rt),
\end{align}
}
\noindent where $-\pi\leq\lambda\leq\pi$, $-\frac{\pi}{2}\leq\theta\leq\frac{\pi}{2}$, and $r_c = 1 - \cos\theta\cos\theta_c\cos(\lambda-\lambda_c) - \sin\theta\sin\theta_c$. For the boundary represented by the $C^{\infty}$ function, we use $x_c=y_c=z_c=0.9$,$a=0.1$,$b=0.2$,$c=0.09$,$A=0.09$, and $\sigma_1=0.2$. For the boundary represented by the $C^3$ function, we use $x_c = y_c = 0.1$, $z_c=0.2$, $a=b=c=0.1$, $B=0.04$, and $\sigma_2 = \frac{16}{25}$. We set $\lambda_c = 0$ and $\theta_c = \frac{\pi}{2}$. Note that since the functions represent curves/surfaces, their first parametric derivatives are the tangent vectors; for a detailed description on generating these and other geometric quantities from the interpolant, see~\cite{SWFKAPNUM2013}.
}
\begin{figure}[h!]
\centering
\subfloat[Approximating a smooth function]
{
	\includegraphics[scale=0.4]{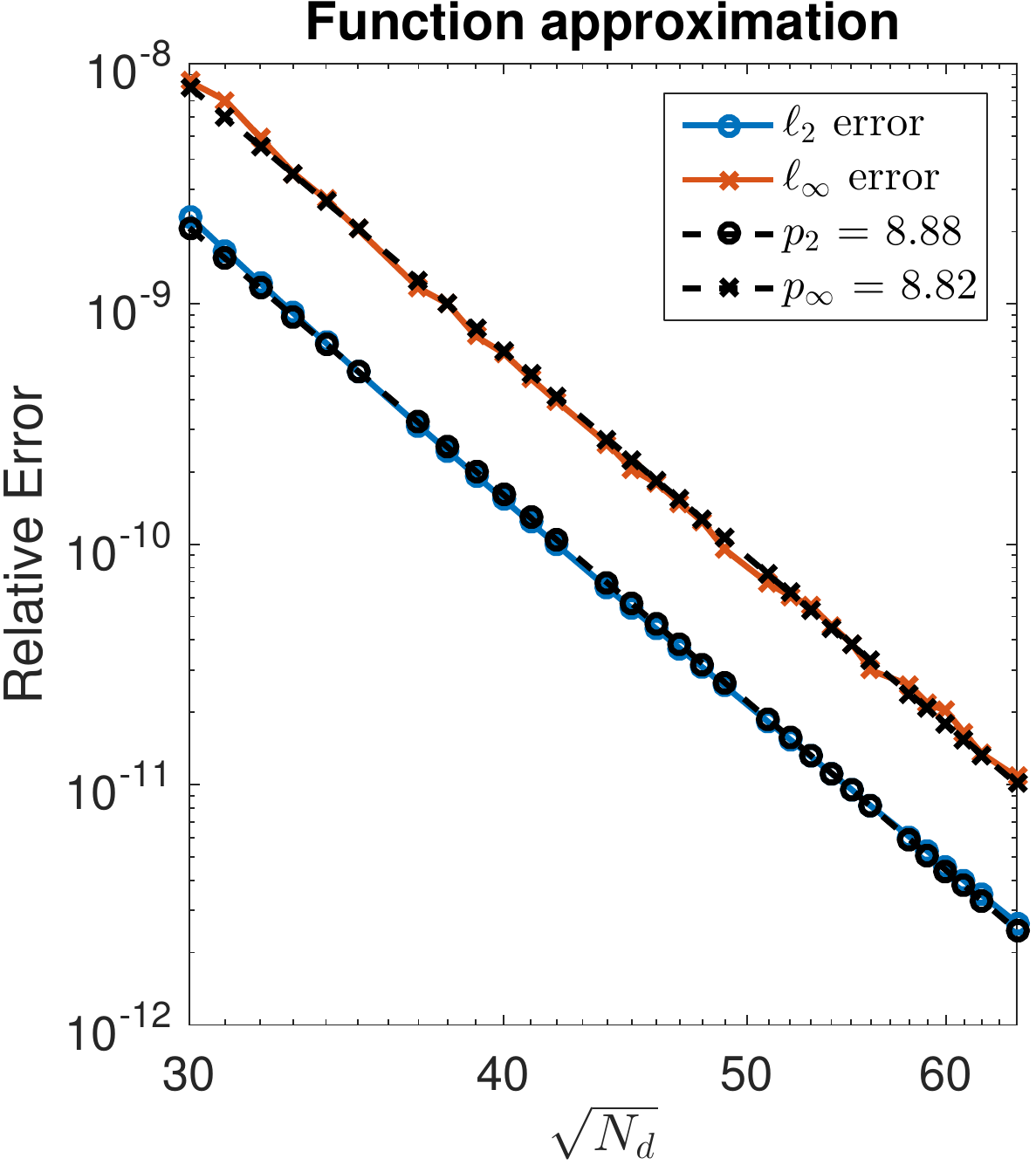} 	
	\label{fig:2a}
}
\subfloat[First derivative with respect to $\lambda^1$]
{
	\includegraphics[scale=0.4]{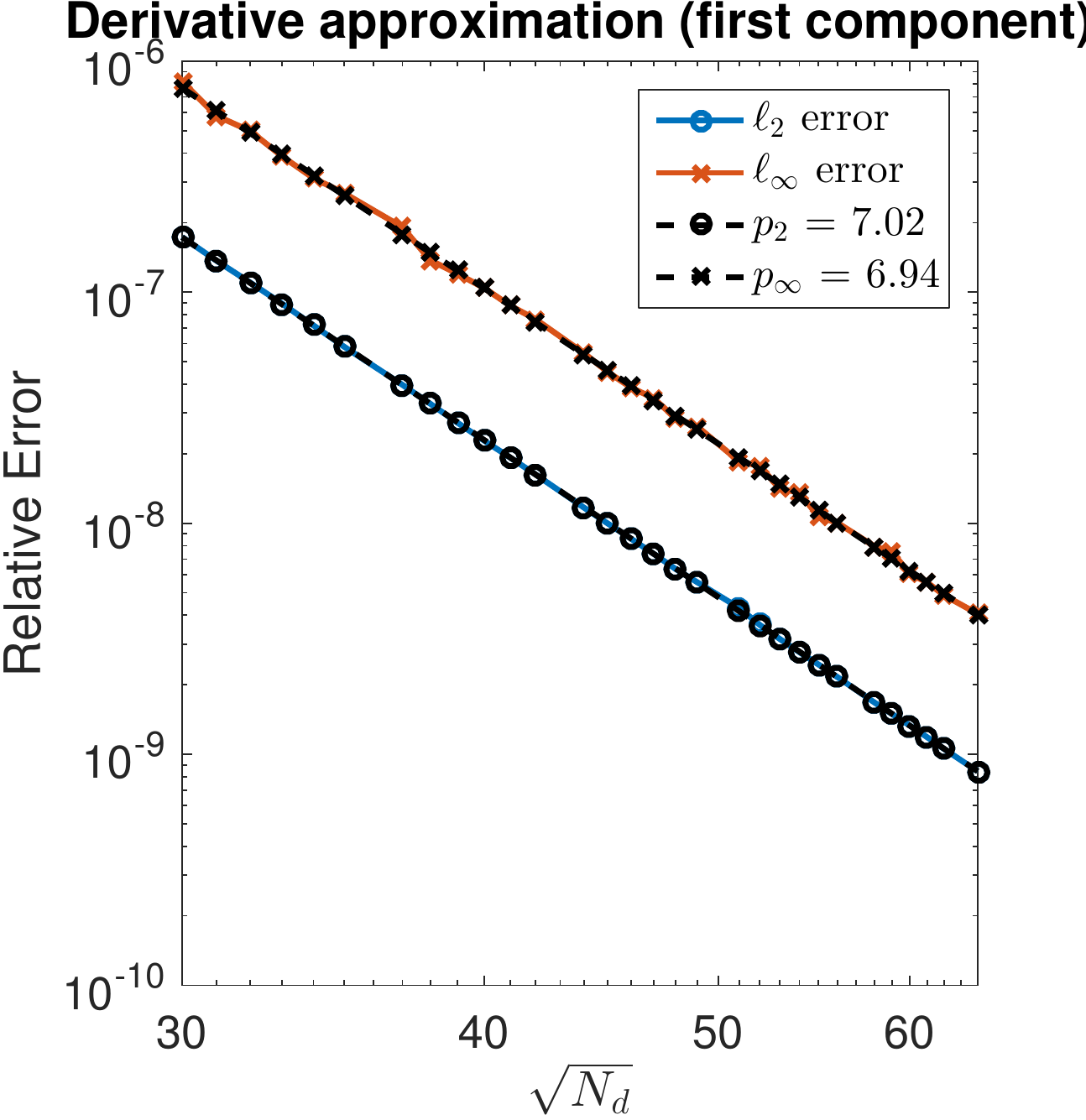}	
	\label{fig:2b}
}

\subfloat[Approximating a $C^3$ function]
{
	\includegraphics[scale=0.4]{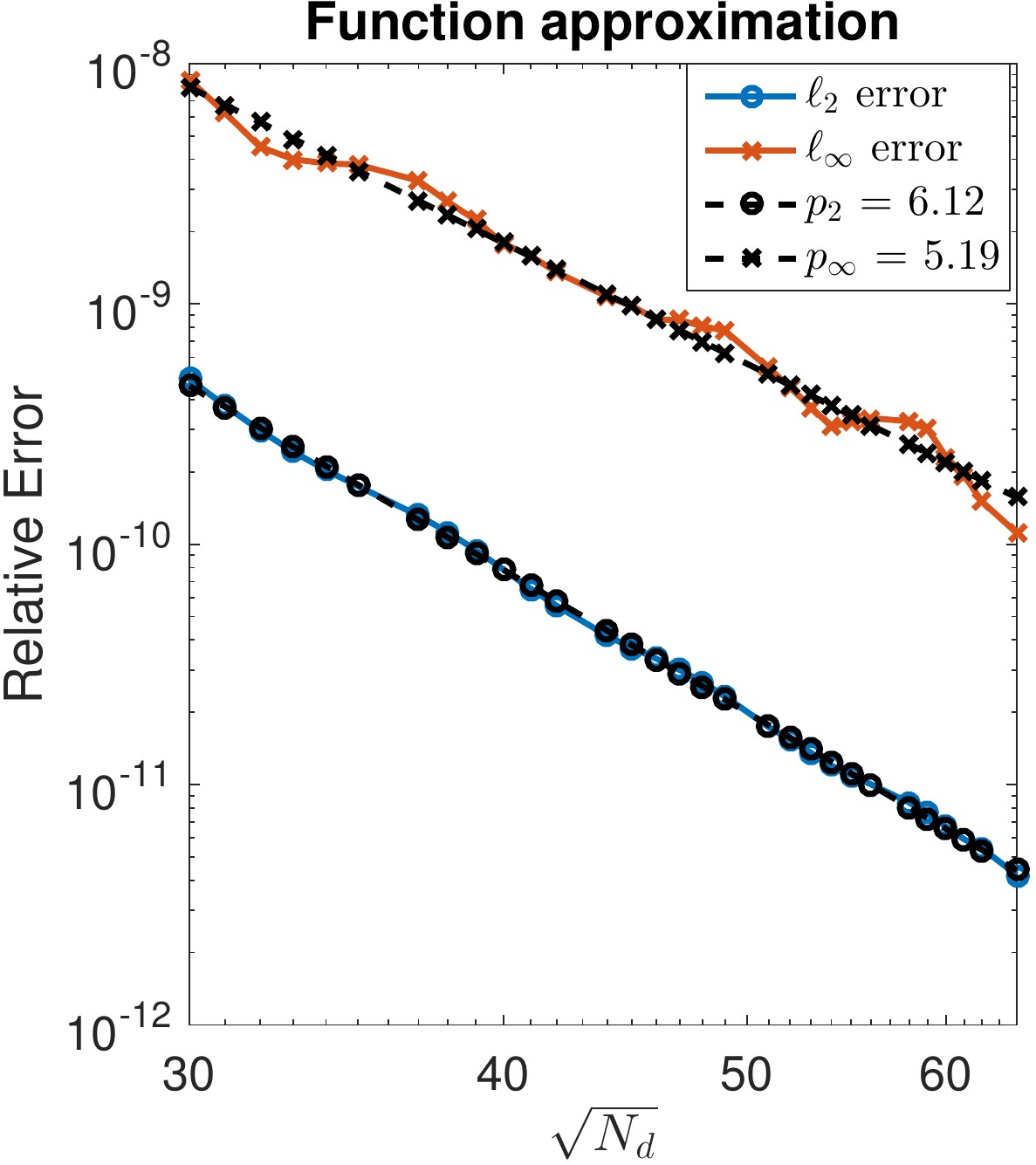} 	
	\label{fig:2c}
}
\subfloat[First derivative with respect to $\lambda^1$]
{
	\includegraphics[scale=0.4]{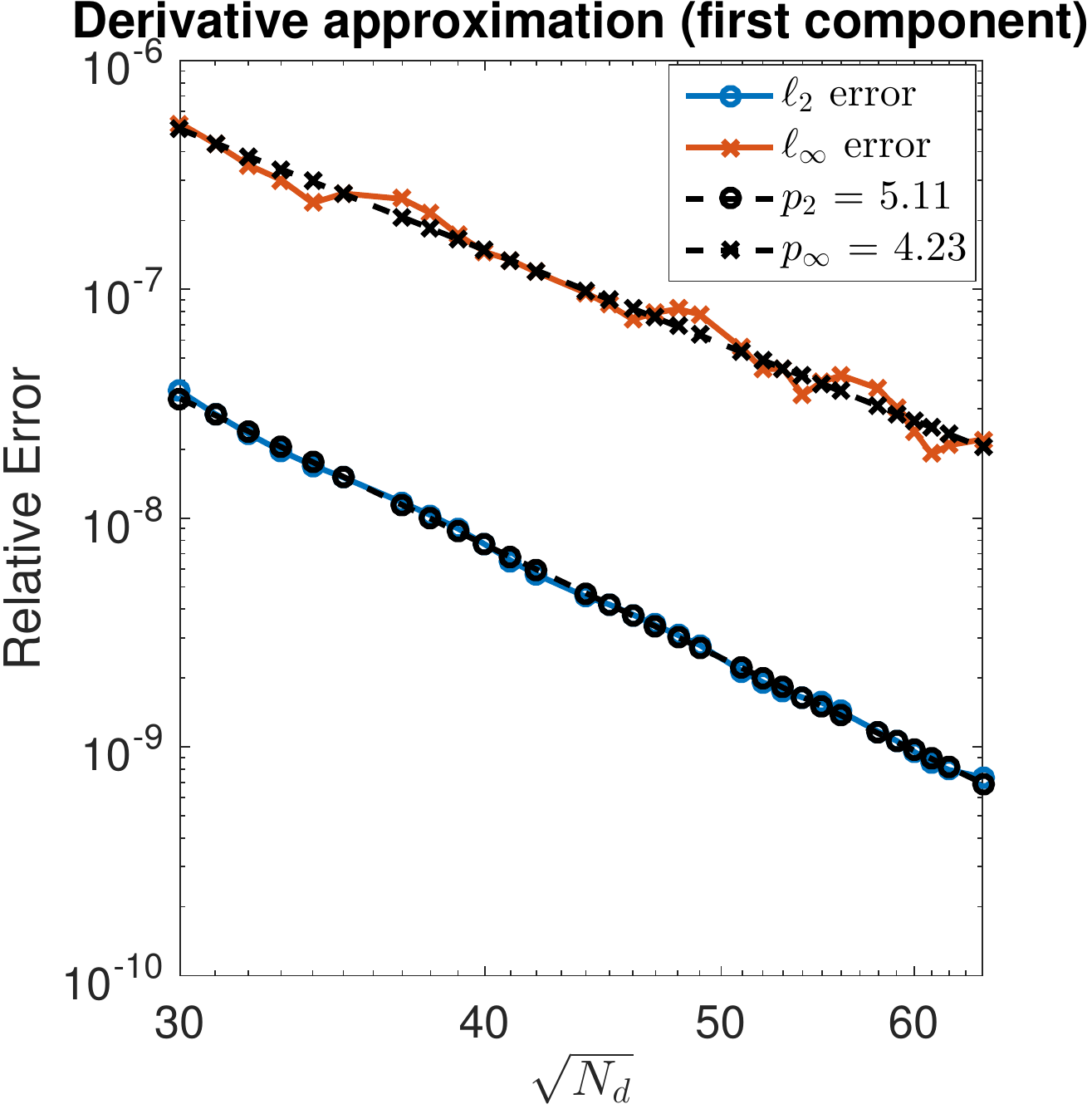} 	
	\label{fig:2d}
}
\caption{Errors in function and derivative approximation for closed 2D surfaces using $\phi(r) = r^6 \log r$. The results for the other first derivative are similar and therefore omitted. The dashed lines in the figures on the bottom row are lines of best fit with slopes shown in the legends.}
\label{fig:2Dresults}	
\end{figure}
We first present results using PHS SBFs for interpolating the above closed 1D curves, then present analogous results for interpolating the above closed 2D surfaces. In the 1D case, we use $\phi(r) = r^7$, while in the 2D case we use $\phi(r) = r^6 \log r$. \comment{Figure \ref{fig:1Dresults} shows the results of modeling 1D curves of differing smoothness with PHS SBFs. Figures \ref{fig:1a} and \ref{fig:1b} show the errors in approximating an infinitely-smooth function and its first derivative using PHS SBFs. Figure \ref{fig:1a} shows that we get a convergence rate of approximately $N_d^{-9}$ ($9$th order) when interpolating the $C^{\infty}$ function with $\phi(r) = r^7$, and lose an order when approximating its derivative. This is an order higher than the theoretical $O(N_d^{-8})$ convergence rate for PHS RBFs in 1D~\cite{Fasshauer:2007}. In contrast, Figures \ref{fig:1c} and \ref{fig:1d} show that when approximating a boundary represented by a $C^2$ function, the convergence rates are limited by the smoothness of the function, with each subsequent derivative converging at one order lower, as indicated by the slopes of the dashed lines. 

Figure \ref{fig:2Dresults} shows similar results for modeling boundaries that are 2D surfaces. Let $h_d \propto \frac{1}{\sqrt{N_d}}$. The predicted convergence rate for the kernel $\phi(r) = r^6 \log r$ is $O(h_d^{8})$ in 2D~\cite{Fasshauer:2007}, and the observed convergence rate in Figure \ref{fig:2a} is slightly higher. Figure \ref{fig:2b} shows that we attain the theoretically-predicted $O(h_d^7)$ convergence rate when approximating the first derivative.  Figure \ref{fig:2c} shows that the approximation order is limited in the case of approximating a $C^3$ function. In the infinity norm, we have lost about three orders of convergence when compared to the $C^{\infty}$ case, and one further order when approximating the derivative.

It is important to note that while spectral convergence rates can be obtained for infinitely-smooth functions using infinitely-smooth SBFs~\cite{SWFKAPNUM2013}, the shape parameters in that work had to be carefully selected by running extensive experiments. In contrast, the PHS SBFs in this study required no tuning to achieve high order convergence rates. Our experiments also indicated that using spherical harmonics in conjunction with SBFs resulted in spectral convergence rates for modeling 2D surfaces. However, we believe that our current formulation strikes an excellent balance between computational cost and accuracy for both $C^{\infty}$ and finitely-smooth boundaries.} \revone{When selecting $N_d$ in practical applications, it is reasonable to select it by ensuring that the error of the geometric model (from theory) is equal to or lower than the error from an RBF-FD discretization.}

%% file: NodeGeneration.tex
\section{Robust node generation on irregular domains and surfaces}
\label{sec:node_gen}
%\begin{figure}[h!]
%\centering
%\subfloat[Boundary nodes on a 3D domain]
%{
	%\includegraphics[scale=0.3]{"Figures/bs_bdry".pdf}
	%
	%\label{fig:3b}
%}
%\subfloat[Interior nodes in a 3D domain]
%{
	%\includegraphics[scale=0.3]{"Figures/bs_int".pdf} 	
	%\label{fig:3c}
%}
%\caption{Node sets generated by Algorithm \ref{alg:node_gen_1} on an irregular 3D domain.}
%\label{fig:node_gen}	
%\end{figure}
Having tested our geometric modeling technique, we present our algorithm (Algorithm \ref{alg:node_gen_1}) for generating scattered nodes on irregular domains $\Omega \subset \mathbb{R}^d$, $d=2,3$ using a combination of Poisson disk sampling and the new geometric modeling technique. Algorithm \ref{alg:node_gen_1} has the important feature that it ensures that interior nodes maintain a user-specified separation distance of $h$ from the boundary and between each other. In the remainder of this section, we will discuss the salient features of Algorithm \ref{alg:node_gen_1}, present some modifications to allow for time-varying embedded boundaries, explore the quality of the node sets it produces (using histograms), and test its suitability for RBF-FD discretizations.
\begin{algorithm}
\caption{Node generation for domains with smooth boundaries}
\label{alg:node_gen_1}
\begin{algorithmic}[1]	
  \Statex{\bf Given}: $\chid = \{(\vX_d)_{\ell}\}_{\ell=1}^{N_d}$, a set of ``seed'' nodes  on domain boundary.
	\Statex{\bf Given}: $h$, the average separation distance between nodes.
	\Statex{\bf Generate}: $\chib = \{(\vX_b)_j\}_{j=1}^{N_b}$, a set of boundary nodes with spacing $h$.
	\Statex{\bf Generate}: $\eta_b$, the set of outward unit normals on the boundary.
	\Statex{\bf Generate}: $\chii = \{\vx_k\}_{k=1}^{N_i}$, a set of interior nodes with spacing $h$.
	\Statex{\bf Generate}: $\rchi = \chii \cup \chib$.
	\State Obtain $\chib$ using Algorithm \ref{alg:ss_decim}.
	\State Evaluate derivatives of \eqref{eq:sbf_poly_model} at the parametric evaluation points obtained from Algorithm \ref{alg:ss_decim} to obtain $\eta_b$.
	\State Use the normals $\eta_b$ to project $\chib$ inwards a distance $h$, giving a set $\hat{\chib}$; this set defines an inner boundary and an inner domain.
	\State Build a kd-tree on the set $\hat{\chib}$.
	\State Generate the oriented bounding volume (OBV) corresponding to $\hat{\chib}$ using Algorithm \ref{alg:pca_obb}.
	\State Fill the OBV with points of (approximate) spacing $h$ using Poisson disk sampling in Algorithm \ref{alg:pds}~\cite{Bridson07}.
	\State Test each sample against the outward normal at its closest inner boundary point (found using the kd-tree). If it is outside the inner domain, discard. 
	\State All remaining Poisson disk samples (not including $\hat{\chib}$) form the set $\chii$.
	\State Set $\rchi = \chii \cup \chib$.
\end{algorithmic}
\end{algorithm}

\subsection{Sampling on surfaces}
\label{sec:node_gen_surf}
A node generation algorithm for RBF-FD methods must generate boundary nodes that are approximately a user-specified distance $h$ apart. Since all domain boundaries are submanifolds of $\mathbb{R}^d$, this problem involves generating a set of parametric evaluation nodes that result in approximately quasi-uniform Cartesian node sets on surfaces. We present an algorithm based on two simple ideas: supersampling of our geometric model, and decimation. As implemented in Algorithm \ref{alg:ss_decim}, we first supersample the parametric interpolant \eqref{eq:sbf_poly_model} to the surface at a large set of parametric evaluation points in $[-\pi,\pi)$ (curves) and $[-\pi,\pi)\times [-\pi/2, \pi/2)$ (surfaces). The resulting Cartesian nodes are not quasi-uniformly spaced, but tend to ``bunch'' according to the parametric map. Thus, the Cartesian node set is now thinned (or \emph{decimated}) to approximately enforce that all nodes be no closer than some separation distance $h$. This requires only the implementation of a ball query (range search) algorithm-- in our case provided by our kd-tree implementation-- in conjunction with a simple depth-first traversal of the node set (in any order). This approach is often referred to as \emph{sample elimination}. A strength of our approach is that the resulting set of samples comes from the geometric model, and retains the high-order accuracy conferred by the model.
\begin{algorithm}
\caption{Sampling on surfaces}
\label{alg:ss_decim}
\begin{algorithmic}[1]
  \Statex{\bf Given}: $\chid = \{(\vX_d)_{\ell}\}_{\ell=1}^{N_d}$, a set of ``seed'' nodes  on domain boundary.
	\Statex{\bf Given}: $h$, the average separation distance between nodes.
	\Statex{\bf Given}: $\tau$, the supersampling parameter.
	\Statex{\bf Generate}: $\chib = \{(\vX_b)_j\}_{j=1}^{N_b}$, a set of boundary nodes  with spacing $h$.
	\Statex{\bf Generate}: $\Lambda^e$, a set of parametric evaluation nodes for \eqref{eq:sbf_poly_model} 
	\Statex{\bf Generate}: the SBF geometric model in \eqref{eq:sbf_poly_model}.
	\State Using \eqref{eq:sbf_poly_model}, fit a geometric model to the nodes in $\chid$.
	\State To find $N_b$ corresponding to a value of $h$, estimate the surface area (or perimeter) $a_d$ of the OBV corresponding to $\chid$ (generated by Algorithm \ref{alg:pca_obb}). Then, $N_b = a_d h^{-(d-1)}$.
	\State Evaluate \eqref{eq:sbf_poly_model} at $\hat{N}_b = \tau N_b$ parametric evaluation points either in $[-\pi,\pi)$ or $[-\pi,\pi)\times [-\pi/2, \pi/2)$. Let $\hat{\chib}$ be the resulting candidate set of Cartesian evaluation points.
	\State Build a kd-tree on $\hat{\chib}$.
	\State Initialize $g$, an $\hat{N}_b$ array of flags, to 1.
	\State Store $\hat{\chib}$ in $\hat{N}_b \times d$ matrix $\hat{X}_b$.
	\For {$k=1,\hat{N}_b$}
		\If {g(k)$ \neq 0$}
				\State Set idxs = indices of points within distance $h$ of $\hat{X}_b$ (using, for instance, a kd-tree ball query).
				\State Set g(idxs $\neq k$) = 0.
		\EndIf
	\EndFor
	\State Set $\hat{g} =$ find(g=1), the vector of indices corresponding to flags of 1.
	\State Collect all parametric evaluation nodes corresponding to indices in $\hat{g}$ into array $\Lambda^e$.
	\State Evaluate t\eqref{eq:sbf_poly_model} at $\Lambda^e$ to obtain $\chib$, the $N_b$ Cartesian points on the boundary.
\end{algorithmic}
\end{algorithm}
Algorithm \ref{alg:ss_decim} clearly only approximately enforces that points be a distance $h$ apart in the \emph{Euclidean} norm, much like Bridson's algorithm for Poisson disk sampling~\cite{Bridson07}. We will explore the spatial distributions of the node sets produced by Algorithm \ref{alg:ss_decim} in Section \ref{sec:res_hist1}. 

\subsection{Oriented bounding boxes (OBBs) from PCA}
\begin{algorithm}
\caption{OBB generation using PCA}
\label{alg:pca_obb}
\begin{algorithmic}[1]	
  \Statex{\bf Given}: $X_b$, the $N_b \times d$ matrix containing a set of points for which we want an OBB.
	\Statex{\bf Generate}: $B$, the matrix of OBB vertices.
	\State Compute $\vx_c = \frac{1}{N_b} \begin{bmatrix} \sum\limits_{i=1}^{N_b} X_b(i,1) & \sum\limits_{i=1}^{N_b} X_b(i,2) & \sum\limits_{i=1}^{N_b} X_b(i,3) \end{bmatrix}$, the centroid of $X_b$.
	\State Compute $M = \mathbf{1} \otimes \vx_c$, the $N_b \times d$ matrix with $\vx_c$ as each of its rows.
	\State Compute $C = \frac{1}{N_b}(X_b-M)^T (X_b-M)$, the $d \times d$ normalized covariance matrix corresponding to $X_b$.
	\State Decompose $C$ as $C =  VDV^T$, where $V$ contains eigenvectors (rotations about origin), $D$ contains eigenvalues (scaling).
	\State Compute unrotated boundary points $\hat{X}_b = X_b V$.
	\State Find column minimum and maximum values for $\hat{X_b}$, \emph{i.e.,} two vertices of the unrotated bounding box.
	\State Using these two vertices, find the side lengths of the box along each coordinate direction.
	\State Using two vertices and side lengths, find the other $2^d - 2$ vertices of the bounding box.
	\State Store all unrotated bounding box vertices in $\hat{B}$, a $2^d \times d$ matrix.
	\State Compute the OBB vertices $B = \hat{B} V^T$.
\end{algorithmic}
\end{algorithm}
Algorithm \ref{alg:node_gen_1} requires an oriented-bounding volume (OBV) for node generation. Almost any simple shape will serve as a bounding volume, but we restrict ourselves to oriented bounding boxes (OBBs), since intersection tests of vectors against boxes simply involve testing a set of linear inequalities. Further, the Poisson disk sampling algorithm is trivial to implement for a rectangular/cuboidal domain. Our goal is to use a technique that has low computational complexity, is easy to implement, and is meshfree. Further, a technique that produces approximately minimal volume OBBs is desirable, since this will result in fewer inside/outside tests. Principal Component Analysis (PCA) fits all our requirements~\cite{Dimitrov2006}, and can easily be used to generate an OBB for our domain in $O(N_b)$ operations. This procedure is summarized in Algorithm \ref{alg:pca_obb}.

\subsection{Poisson disk sampling}

\comment{Algorithm \ref{alg:node_gen_1} is primarily structured around Poisson disk sampling inside the OBB generated by Algorithm \ref{alg:pca_obb}. Our approach for Poisson disk sampling in the OBB is a straightforward implementation of the algorithm described in~\cite{Bridson07}. For completeness, we present this procedure in Algorithm \ref{alg:pds}, adapted to our notation.
\begin{algorithm}
\caption{Fast Poisson disk sampling}
\label{alg:pds}
\begin{algorithmic}[1]	
  \Statex{\bf Given}: $h$, the minimum distance between nodes.
	\Statex{\bf Given}: $d$, the spatial dimension.
	\Statex{\bf Given}: $\hat{k}$, the number of samples to choose before rejection (the Poisson neighborhood size).
	\Statex{\bf Given}: $B$, the matrix of vertices of the domain OBB.
	\Statex{\bf Generate}: The set of $N_{obb}$ Poisson disk samples in the domain OBB.
	\State Initialize a $d$-dimensional Cartesian background grid $\calG$ with cell size $\frac{h}{\sqrt{d}}$. A value of $-1$ in $\calG$ indicates no sample, any non-negative integer is the \emph{linear index} of the sample in a cell.
	\State Generate the first uniform random node $\vx_0$ within the domain OBB. Insert this node into $\calG$.
	\State Initialize the array of node indices (active list) $\calI$ with the index of $0$ (corresponding to $\vx_0$).
	\While{$\calI$ is not empty}
	\State Choose a random index $i$ from $\calI$.
	\State Generate $\hat{k}$ uniform random nodes in the spherical annulus centered at $\vx_i$ with inner radius $h$ and outer radius $2h$. Place these nodes in set $\mathcal{P}$.
	\State Compare all of $\mathcal{P}$ against existing nodes within distance $h$ (using $\calG$ to facilitate comparisons).
	\State If none of the nodes in $\mathcal{P}$ are sufficiently far from existing samples, remove $i$ from the active list.
	\State If any of the nodes in $\mathcal{P}$ are sufficiently far from existing samples, add their indices to the active list, store these \emph{valid samples} in the set $\mathcal{V}$.	
	\State If any of the nodes in $\mathcal{P}$ are outside the domain OBB, reject them.
	\EndWhile
	\State The set of $N_{obb}$ Poisson disk samples is $\mathcal{V}$.
\end{algorithmic}
\end{algorithm}
The complexity of this algorithm is $O(\hat{k} N_{obb})$. Since $\hat{k}$ is a constant, the cost scales as $O(N_{obb})$. In Section \ref{sec:results}, we will explore the impact of $\hat{k}$ on the uniformity of node distributions and on the accuracy for RBF-FD discretizations of PDEs. We explore the node distributions obtained from the use of Algorithm \ref{alg:pds} within Algorithm \ref{alg:node_gen_1} in Section \ref{sec:res_hist2}.}

\subsection{Domains with time-varying embedded boundaries}

Biological problems often involve domains with not only irregular outer boundaries, but irregular embedded inner boundaries, \emph{e.g.,} platelets and red blood cells in a blood vessel~\cite{SWFKIJNMF2015}. While the authors have begun developing numerical methods for simulations in such complex domains~\cite{SWFKIJNMF2014,ShankarJCP2017,SFJCP2018}, our methods currently lack a robust node generation algorithm that can modify its node sets \emph{locally} when the embedded inner boundaries deform, and are added or removed.
\begin{figure}[h!]
\centering
\subfloat[Star domain]
{
	\includegraphics[scale=0.3]{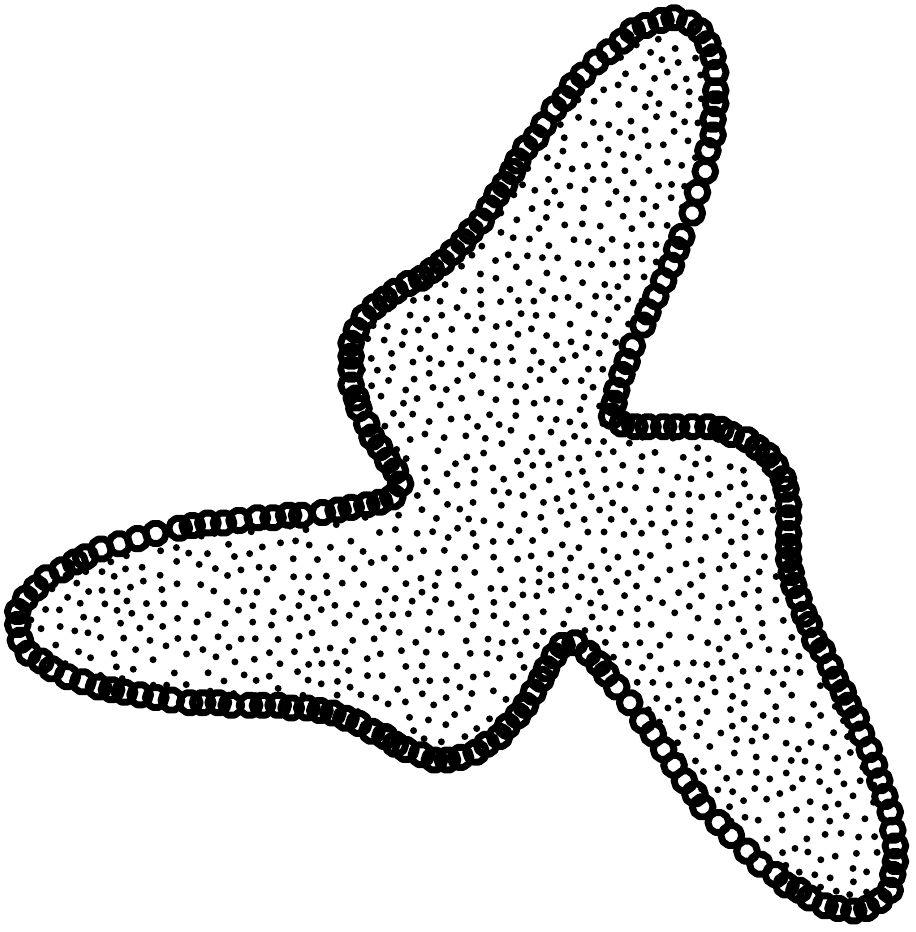} 	
	\label{fig:ea}
}
\subfloat[Star with embedded ellipse]
{
	\includegraphics[scale=0.3]{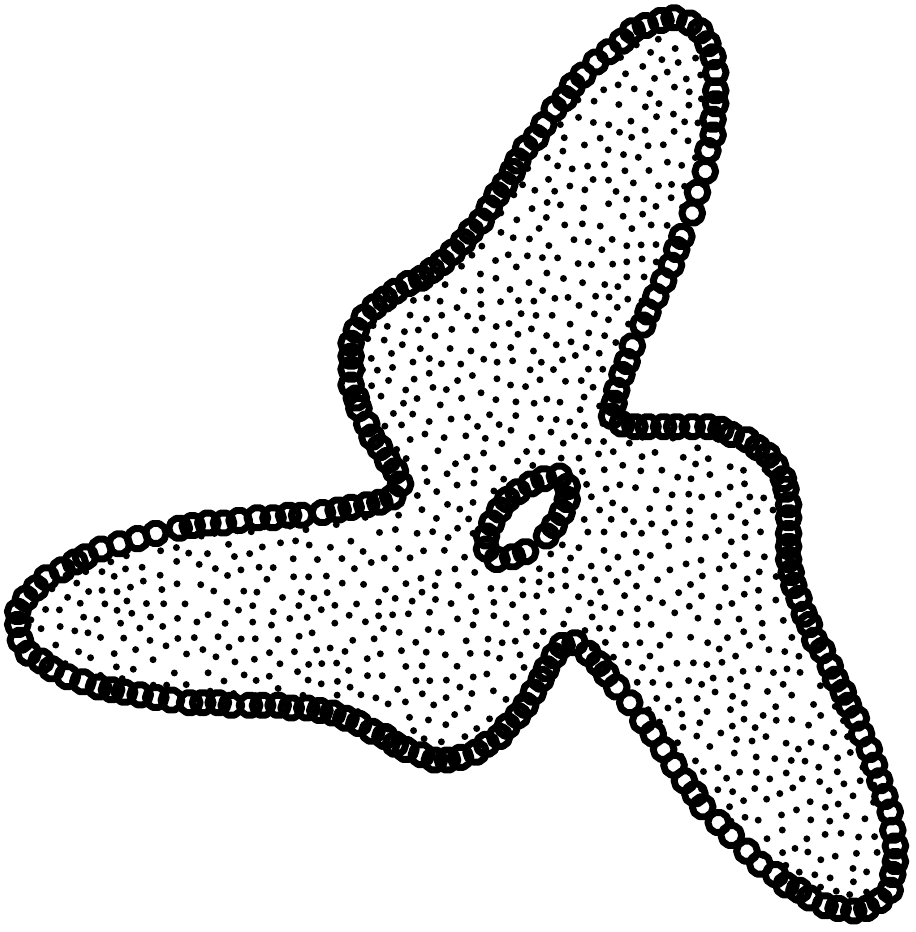}
	
	\label{fig:eb}
}
\caption{Local node modification for embedded boundaries using Algorithm \ref{alg:node_mod}. The embedded ellipse is shown using circles.}
\label{fig:node_mod}
\end{figure}
Fortunately, Algorithm \ref{alg:node_gen_1} is easily modified to tackle this problem. Consider the scenario where Algorithm \ref{alg:node_gen_1} has generated node sets $\chib$ and $\chii$ on the boundary and interior of an irregular domain $\Omega_0$ without any embedded boundaries. Now consider the inclusion of irregular boundaries $\Gamma = \{\Gamma_j\}_{j=1}^{N_p}$ enclosing domains $\{\Omega_j\}_{j=1}^{N_p}$ into $\Omega_0$ in such a way that these boundaries define a new domain $\Omega = \Omega_0 \backslash \bigcup\limits_{j=1}^{N_p}\Omega_j$. Our node generator must automatically and efficiently generate a node set in $\Omega$ that \emph{deletes} the nodes contained in $\tilde{\Omega} = \bigcup\limits_{j=1}^{N_p}\Omega_j$. In fact, our algorithm also deletes any nodes that are within an $h$ distance of the embedded boundaries, again ensuring that the global separation distance is approximately $h$. This is detailed in Algorithm \ref{alg:node_mod}.
\begin{algorithm}
\caption{Node set modification for (closed) embedded boundaries}
\label{alg:node_mod}
\begin{algorithmic}[1]	
	\Statex{\bf Given}: $h$, the average separation distance between nodes in domain $\Omega_0$.
	\Statex{\bf Given}: $\alpha$, the ratio of the numbers of inner boundary nodes to outer boundary nodes (see Section \ref{sec:nodemod_complex}).
	\Statex{\bf Given}: $\chib = \{(\vX_b)_j\}_{j=1}^{N_b}$, a set of boundary nodes on $\partial \Omega_0$.
	\Statex{\bf Given}: $\eta_b$, the set of outward unit normals on the boundary $\partial \Omega_0$.
	\Statex{\bf Given}: $\rchi = \{\vx_k\}_{k=1}^{N}$, the set of nodes on $\Omega_0$ (including the boundary).	
	\Statex{\bf Given}: $N_d^{\Gamma}$ seed nodes on each of the embedded boundaries $\Gamma_j, j=1,\hdots,N_p$.	
	\Statex{\bf Generate}: $Z$, the $N \times 2$ array whose first column indicates whether a point $\vx_k \in \Omega$, and second column is $j$ if $\vx_k \in \Omega_j$.
	\Statex{\bf Generate}: $\tilde{\rchi}$, the (modified) set of $\tilde{N}$ nodes on the irregular domain $\Omega = \Omega_0 \backslash \tilde{\Omega}$.	
	\State Using \eqref{eq:sbf_poly_model}, fit a geometric model to the seed nodes on each embedded boundary $\Gamma_j,j=1,\hdots,p$.
	\State Set $N_b^{\Gamma} = \alpha N_b$.
	\State For $j=1,\hdots,N_p$, obtain $\chibgj$, the $N_b^{\Gamma}$ boundary points on the $j$th embedded boundary using Algorithm \ref{alg:ss_decim}.
	\State Evaluate derivatives of the $N_p$ boundary interpolants at $N_b^{\Gamma}$ parametric points to obtain $\eta_b^{\Gamma_j}$, the set of unit normals on each boundary (pointing into $\Omega_0$).	
	\State Extend each embedded boundary $\Gamma_j$ by distance $h$ in their normal directions to obtain $\tilde{\Gamma}_j$ and the corresponding points $\tilde{\chibgj}$.
	\State Let $\tilde{\chibg} = \bigcup\limits_{j=1}^{N_p} \tilde{\chibgj}$.
	\State Add $\tilde{\chibg}$ to the domain kd-tree currently containing $\chib$.
	\State Generate the OBB for each $\tilde{\Gamma}_j$ using Algorithm \ref{alg:pca_obb} on the $N_b^{\Gamma}$ boundary points.	
	\State Initialize the 2D array (map) Z with dimensions $N \times 2$ to zeros.
	\For {$k=1,N$}
		\For{$j=1,N_p$}
				\If {$\vx_k$ is within the $j$\textsuperscript{th} OBB}
					\State Find the closest point $\zeta$ on $\tilde{\Gamma}_j$ using the domain kd-tree.
					\State Use $\zeta$ to test $\vx_k$ against $\tilde{\Gamma}_j$.
					\If {$\vx_k$ is inside $\tilde{\Gamma}_j$}
						\State Set $Z(k,1)$ = 1, $Z(k,2) = j$.
						\State Break and go to next $k$ value.
					\EndIf
				\EndIf				
		\EndFor
	\EndFor				
	\State Let g = find(Z(:,1)=0 be the list of indices of points with $\Omega$. 
	\State Set $\tilde{\rchi} = \rchi(g=0)$, $\tilde{N}$ = length($\tilde{\rchi}$).
\end{algorithmic}
\end{algorithm}
This procedure involves only locally changing the original node set as new objects are introduced, and also accounts for cases where an embedded boundary may touch the domain outer boundary. It is important to note that this algorithm features potential early termination: a point is only tested against bounding boxes and embedded boundaries until it is found to be contained within an embedded boundary. While other efficiency improvements are possible, we forgo them for simplicity. Figure \ref{fig:node_mod} shows an example of the node sets obtained using Algorithm \ref{alg:node_mod}. In Figure \ref{fig:ea}, we show the interior nodes on the star domain. Figure \ref{fig:eb} shows the node set obtained when an ellipse is embedded into the star domain. It is easy to see that Algorithm \ref{alg:node_mod} only locally modifies node sets, as desired.

Algorithm \ref{alg:node_mod} does not take into account the case where an embedded boundary is removed. There are two options within the above framework for modifying the node sets in such a situation. The first option is to fully generate a new node set on $\Omega$ after an embedded boundary is removed; this would involve using Algorithm \ref{alg:node_mod} to modify the original node set on $\Omega_0$ to account for the embedded boundaries that were \emph{not removed}. This option is rather wasteful. Instead, Algorithm \ref{alg:node_mod} associates each grid point $\vx_k$ with an embedded boundary $\Omega_j$. Thus, to handle a \emph{removed} embedded boundary $\Omega_j$, we need only find all nodes in the original node set that are contained within $\Omega_j$ by searching the second dimension of the array $Z$ for the value $j$. Of course, this second option relies on knowing which of the embedded boundaries were removed. If this is not known, the first option discussed above is preferable. 
\subsection{Boundary refinement and ghost nodes}

The numerical solution of PDEs with RBF-FD sometimes requires denser node sets near domain boundaries~\cite{FlyerNS,FlyerPHS,ShankarJCP2017}, and/or ghost nodes outside the domain boundary to enforce boundary conditions~\cite{FlyerNS,BarnettPHS,SFJCP2018}. Our node generator must therefore possess these capabilities. Generating ghost nodes is straightforward: given a set of boundary nodes and unit outward normals, simply copy the nodes a distance $h$ outside the domain along the outward normals. Further, Algorithm \ref{alg:node_gen_1} can be easily modified to provide a node set refined near a boundary: simply copy the boundary nodes at some user-specified distance inward from the boundary. This places a layer of nodes between $\chib$ and the inner boundary $\hat{\chib}$. This approach generalizes straightforwardly to multiple layers of boundary refinement. However, a drawback of this approach is that it does not generate \emph{graded} boundary-refined node sets,~\emph{i.e.}, node sets that smoothly vary in space. We leave this generalization for future work.

\input{Results}

\section{Complexity Analysis}
\label{sec:complex}

We now derive the computational complexity of the Algorithms \ref{alg:node_gen_1} and \ref{alg:node_mod}.

\subsection{Complexity of one-time node generation}
\label{sec:nodegen_complex}

The total computational cost of our node generator is a sum of the individual costs of the following steps:
\begin{itemize}
\item Forming a kd-tree on the domain boundary: The kd-tree is formed on $N_b$ points for a cost of $C_1 = O(d N_b \log N_b)$. Each subsequent nearest neighbor search costs $O(\log N_b)$ operations.

\item SBF boundary representation: Since this involves $N_d$ seed nodes, it requires the solution of the $N_d \times N_d$ dense linear system \eqref{eq:sbf_linsys}, which has an asymptotic cost of $O(N_d^3)$. This interpolant is evaluated to obtain $\tau N_b$ boundary points for a cost of $O(N_b N_d)$, then thinned for a worst-case cost of $O(N_b \log N_b)$ using the kd-tree. If we assume $N_d = \gamma N_b$, where $\gamma \in (0,1]$ is some small number, the evaluation cost can be written as $O(\gamma N_b^2)$. Cost: $C_2 = O(\gamma^3 N_b^3) + O(\gamma N_b^2) + O(N_b \log N_b)$.

\item OBB: The primary cost in the PCA computation of the OBB is the $O(d^2 N_b)$ cost of forming the $d \times d$ covariance matrix. The subsequent eigendecomposition of the $d \times d$ matrix is computed in $d^3$ flops. Then, the rotations and max/min computations to find the unrotated bounding box cost $O(N_b)$. The costs of rotations of the bounding boxes can be neglected when $d=2,3$. Cost: $C_3 = O(N_b)$.

\item Poisson disk sampling in the OBB: Using Algorithm \ref{alg:pds}, Poisson disk samples for the OBB can be computed for a cost of $C_4 = O(N_{obb})$.

\item Eliminate OBB nodes outside the domain boundary: This requires finding the closest boundary point to each of the $N_{obb}$ Poisson disk samples in the OBB using the kd-tree. The total cost is therefore given by $C_5 = O(N_{obb} \log N_b)$.
\end{itemize}

The total cost is given by $C_{tot} = \sum\limits_{i=1}^5 C_i$. Dropping the big-$O$ notation for convenience, we have
\begin{align}
C_{tot} = d N_b \log N_b + \gamma^3 N_b^3 + \gamma N_b^2 + N_b \log N_b + N_b + N_{obb} + N_{obb} \log N_b.
\label{eq:complexity}
\end{align}
All our costs are currently in terms of $N_b$ and $N_{obb}$. However, it is more appropriate to express $C_{tot}$ in terms of the total number of points in the domain $N = N_i + N_b$. This requires dimension-specific simplifications.

\subsubsection{2D complexity estimates}

We now present the complete 2D complexity estimates in terms of $N$. Assuming the nodes are approximately uniformly spaced on both the boundary and the interior, we have
\begin{align}
h_i = \lf(\frac{a}{N_i}\rt)^{\frac{1}{2}}, h_b = \frac{q}{N_b}.
\end{align}
If we wish the nodes to be spaced comparably on the boundary and interior, it is reasonable to assume that $h_i = h_b = h$. Solving for $N_b$ then gives
\begin{align}
N_b &= q a^{-\frac{1}{2}} N_i^{\frac{1}{2}}.\label{eq:Nb2d}
\end{align}
Substituting \eqref{eq:Nb2d} in \eqref{eq:complexity} and neglecting sublinear terms in $N_i$ gives us
\begin{align}
C_{tot} = \gamma^3 q^3 a^{-\frac{3}{2}} N_i^{\frac{3}{2}} + \gamma \frac{q^2}{a} N_i + 2q a^{-\frac{1}{2}} N_i^{\frac{1}{2}} + 2N_{obb}. 
\label{eq:ctot1}
\end{align}
We must now relate $N_{obb}$ to $N_i$ to get the total cost in terms of $N_i$. We know that
\begin{align}
h_i &= \lf(\frac{a}{N_i}\rt)^{\frac{1}{2}} = \lf(\frac{a_{obb}}{N_{obb}}\rt)^{\frac{1}{2}}, \\
\implies N_{obb} &= \frac{a_{obb}}{a} N_i.
\end{align}
Substituting this new expression into \eqref{eq:ctot1} and retaining only linear and higher-order terms in $N_i$, we obtain
\begin{align}
C_{tot} = \gamma^3 q^3 a^{-1.5} N_i^{1.5} + \gamma \frac{q^2}{a} N_i + 2\frac{a_{obb}}{a} N_i. 
\end{align}
Since $N = N_i + N_b$ and $N_b << N_i$, it is reasonable to replace $N_i$ by $N$, giving us
\begin{align}
C_{tot} = \gamma^3 q^3 a^{-1.5} N^{1.5} + \gamma \frac{q^2}{a} N + 2\frac{a_{obb}}{a} N. 
\end{align}
The constant in front of the leading-order $N^{1.5}$ term is typically small in practice for large $N$ and smooth domain boundaries. In such a case, the $O(N)$ terms dominate. Thus, for small $\gamma$ and reasonably compact domains, we have
\begin{align}
C_{tot} \approx \lf(\gamma \frac{q^2}{a} + 2\frac{a_{obb}}{a}\rt) N.
\end{align}
Our node generator thus has a theoretical computational complexity of $O(N)$ in 2D, despite the use of a global RBF interpolant on the boundary. As we show in Section \ref{sec:scale_res}, we recover the $O(N)$ complexity in practice.

\subsubsection{3D complexity estimates}

We proceed much in the same way as in the 2D case, but obtain a different estimate. First, assuming the nodes are approximately uniformly spaced on both the boundary and the interior, we now have
\begin{align}
h_i &= \lf(\frac{v}{N_i} \rt)^{\frac{1}{3}}, h_b = \lf(\frac{s}{N_b} \rt)^{\frac{1}{2}}.
\end{align}
Now, letting $h_i = h_b = h$, solving for $N_b$ gives
\begin{align}
N_b & = s v^{-\frac{2}{3}} N_i^{\frac{2}{3}}. \label{eq:Nb3d}
\end{align}
Using \eqref{eq:Nb3d} in \eqref{eq:complexity} and neglecting sublinear terms in $N_i$, we have
\begin{align}
C_{tot} = \gamma^3 s^3 v^{-2} N_i^2 + \gamma s^2 v^{-\frac{4}{3}} N_i^{\frac{4}{3}} + 2 N_{obb}.
\end{align}
Once again relating $N_{obb}$ to $N_i$ using the volume of the OBB $v_{obb}$, we have
\begin{align}
C_{tot} = \gamma^3 s^3 v^{-2} N_i^2 + \gamma s^2 v^{-\frac{4}{3}} N_i^{\frac{4}{3}} + 2 \frac{v_{obb}}{v} N_i.
\end{align}
For a sufficiently large number of $N_b$ (small $h$) and fixed $N_d$, $\gamma$ can be as low as $O(10^{-3})$. Thus, both the quadratic and superlinear terms in the above expansion may be small for a compact domain, leaving us with
\begin{align}
C_{tot} \approx 2 \frac{v_{obb}}{v} N_i \approx 2 \frac{v_{obb}}{v} N.
\end{align}
Even if the superlinear term is retained, this represents a complexity of $O(N^{1.\bar{33}})$. However, in practice, we observe close to $O(N)$ complexity, as we show in Section \ref{sec:scale_res}.

\subsection{Complexity of node set modification}
\label{sec:nodemod_complex}

We now derive the complexity associated with Algorithm \ref{alg:node_mod}. The total \emph{worst-case} computational cost for removing nodes from the interior of $N_p$ embedded boundaries can be written as
\begin{align}
C_{mod} = O( N_p(N_d^{\Gamma})^3) + O(N_p N_b^{\Gamma} (N_d^{\Gamma})^2) + O(N_p N_b^{\Gamma}) + O(N N_p),
\end{align}
where the first term corresponds to fitting the SBF geometric model to each embedded boundary, the second term to evaluations of the models to obtain normals, the third term to the computation of the OBBs of the embedded boundaries, and the last term to testing each of the $N$ domain nodes against the $N_p$ OBBs. Now, let
\begin{align}
N_b^{\Gamma} = \alpha N_b, N_d^{\Gamma} = \alpha N_d.
\end{align}
If approximately the same spacing between nodes is maintained on all boundaries, we have $h_{\Gamma} = h_b$, which in 2D gives us $\alpha = \frac{q^{\Gamma}}{q}$ (rounded to the nearest integer). Similarly, in 3D, we have $\alpha = \frac{s^{\Gamma}}{s}$. Dropping the big-$O$ notation for convenience, $C_{mod}$ can be written as
\begin{align}
C_{mod} = \alpha\gamma^2(\alpha^2\gamma + 1) N_p N_b^3 + \alpha N_p N_b + N N_p.
\label{eq:cmod}
\end{align}
Using the fact that $N = N_i + N_b$, \eqref{eq:cmod} can be written as
\begin{align}
C_{mod}^{2\rm D} = \alpha\gamma^2(\alpha^2\gamma + 1)  N_p q^3 a^{-1.5} N_i^{1.5} + \alpha N_p q a^{-\frac{1}{2}} N^{\frac{1}{2}} + N N_p.
\end{align}
By a similar argument as in Section \ref{sec:nodegen_complex}, the $N_i^{1.5}$ term is small, leaving an estimate of
\begin{align}
C_{mod}^{2\rm D} \approx N N_p,
\end{align}
where sublinear terms in $N$ and $N_i$ have been neglected. Since the number of embedded boundaries is always much smaller than the number of nodes, \emph{i.e.,} $N_p << N$, we have
$C_{mod}^{2\rm D} = O(N)$. The 3D derivation is similar. Using the definition of $N_b$ in terms of $N_i$ in 3D, \eqref{eq:cmod} can be re-written as
\begin{align}
C_{mod}^{3\rm D} =  \alpha\gamma^2(\alpha^2\gamma + 1) N_p s^3 v^{-2} N_i^{2} + \alpha N_p s v^{-\frac{2}{3}} N^{\frac{2}{3}} + N N_p.
\end{align}
Again, in practice, the linear term in $N$ dominates.  Thus, we have
\begin{align}
C_{mod}^{3\rm D} \approx N N_p,
\end{align}
which is in practice $O(N)$ since $N_p << N$. If a bounding box hierarchy is used to keep track of the embedded boundaries, the $O(N N_p)$ term can be further shrunk to $O(N \log N_p)$, but we leave this approach for future work. It is important to note that our estimates for node generation only constitute the worst-case. In general, it is unlikely that a Poisson disk sample needs to be tested against \emph{all} OBBs of the embedded inner boundaries. Indeed, the disk sample is deleted as soon as it is found to lie within any embedded boundary, requiring no additional tests. The true complexity estimate in this scenario is non-deterministic. We leave this analysis for future work also.

\subsection{Scaling of the node generator}
\label{sec:scale_res}
\begin{figure}[h!]
\centering
\subfloat[Star, $N_d = 128$]
{
	\includegraphics[scale=0.5]{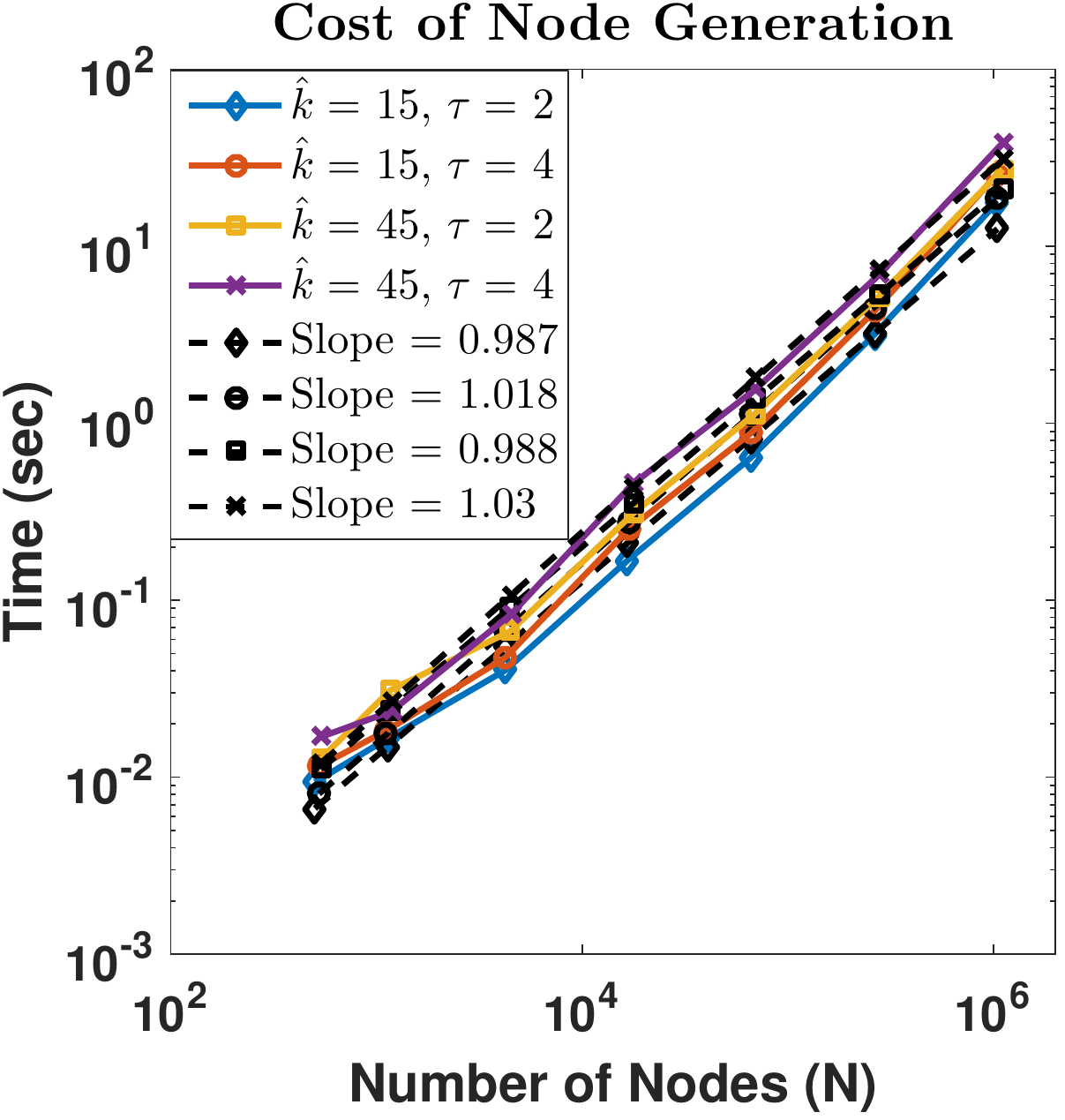}
	
	\label{fig:s1a}
}
\subfloat[Star, $N_d = 256$]
{
	\includegraphics[scale=0.5]{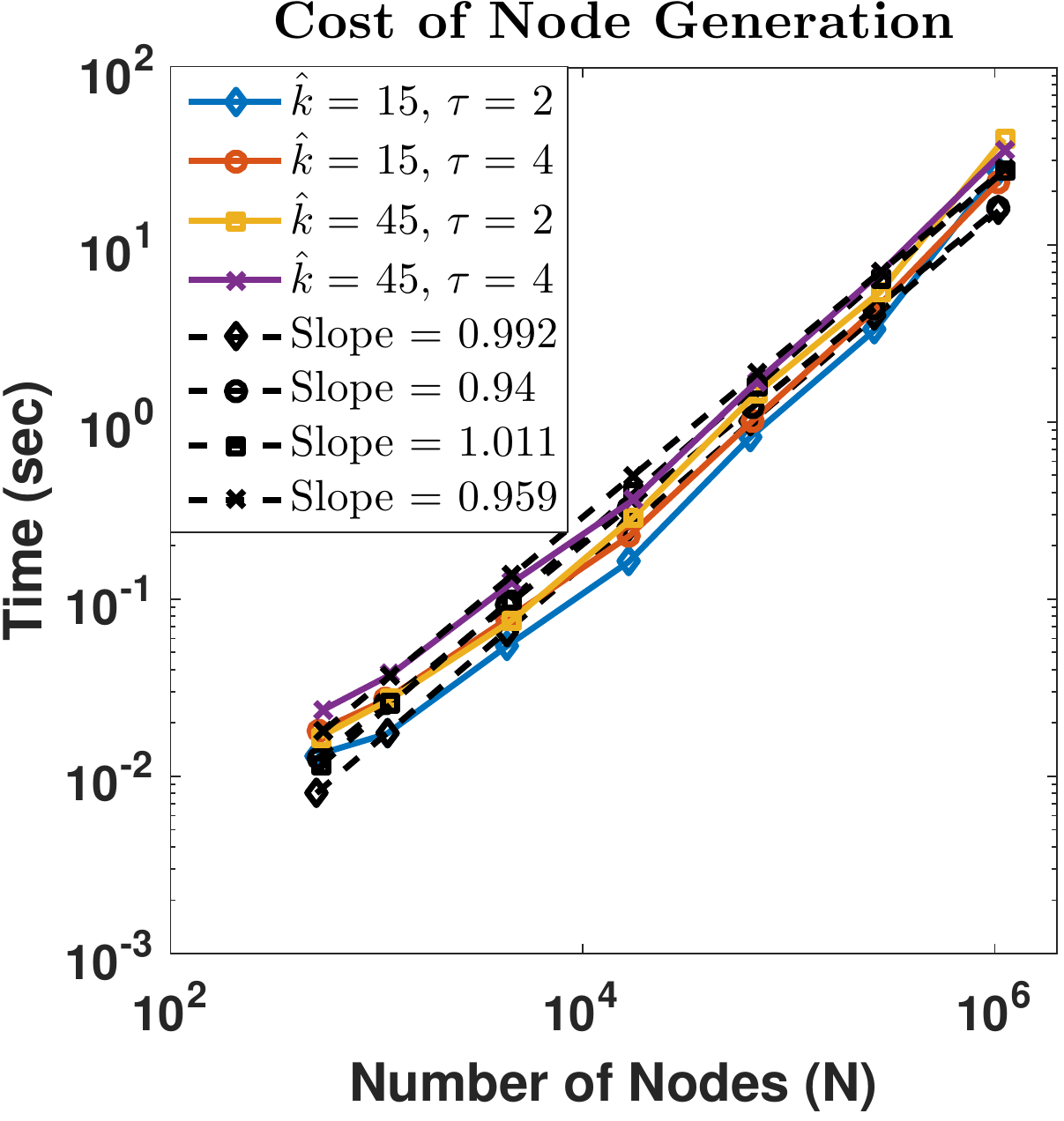} 	
	\label{fig:s1b}
}

\subfloat[Bumpy sphere, $N_d = 400$]
{
	\includegraphics[scale=0.5]{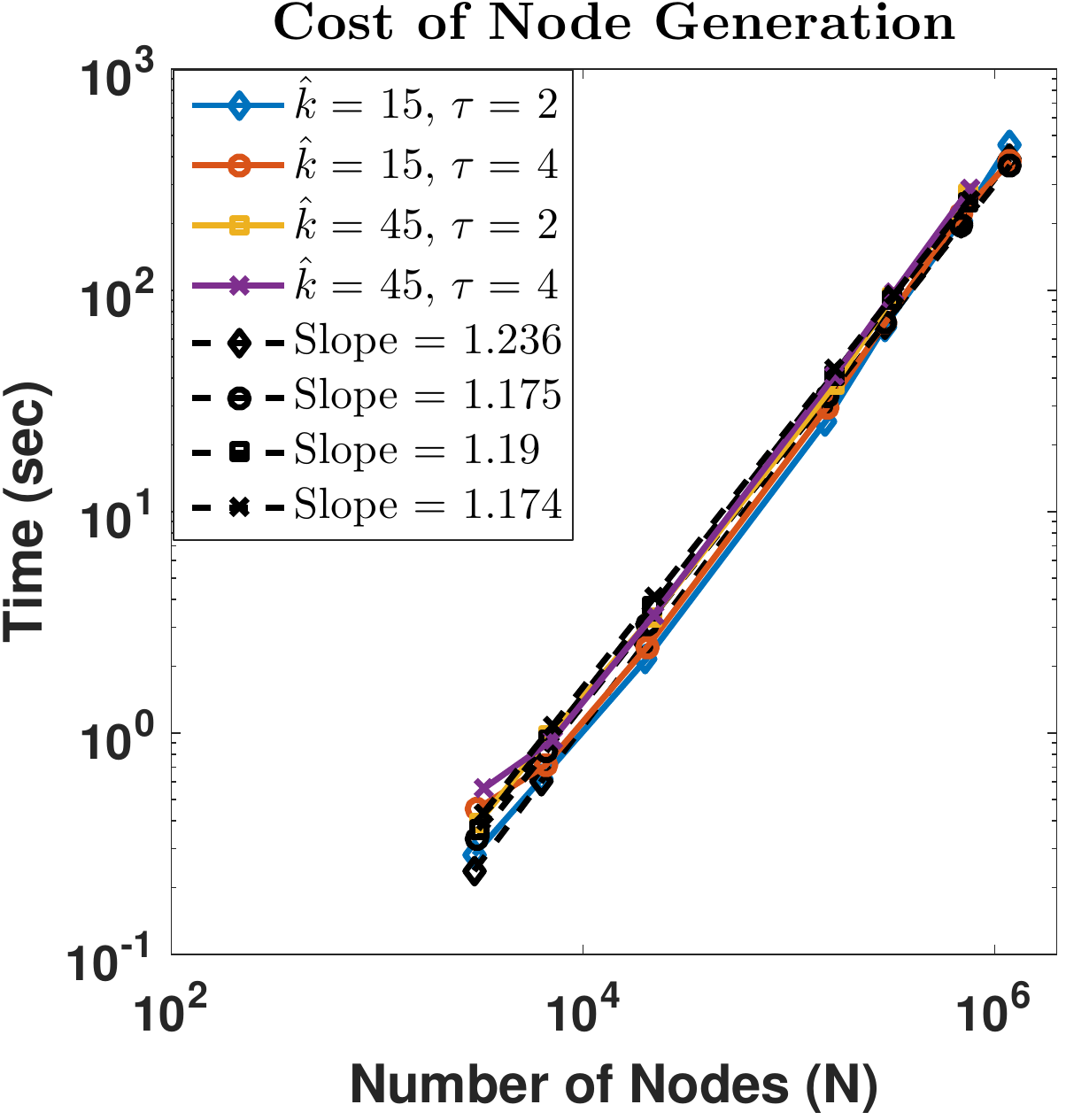}
	
	\label{fig:s1c}
}
\subfloat[Bumpy sphere, $N_d = 800$]
{
	\includegraphics[scale=0.5]{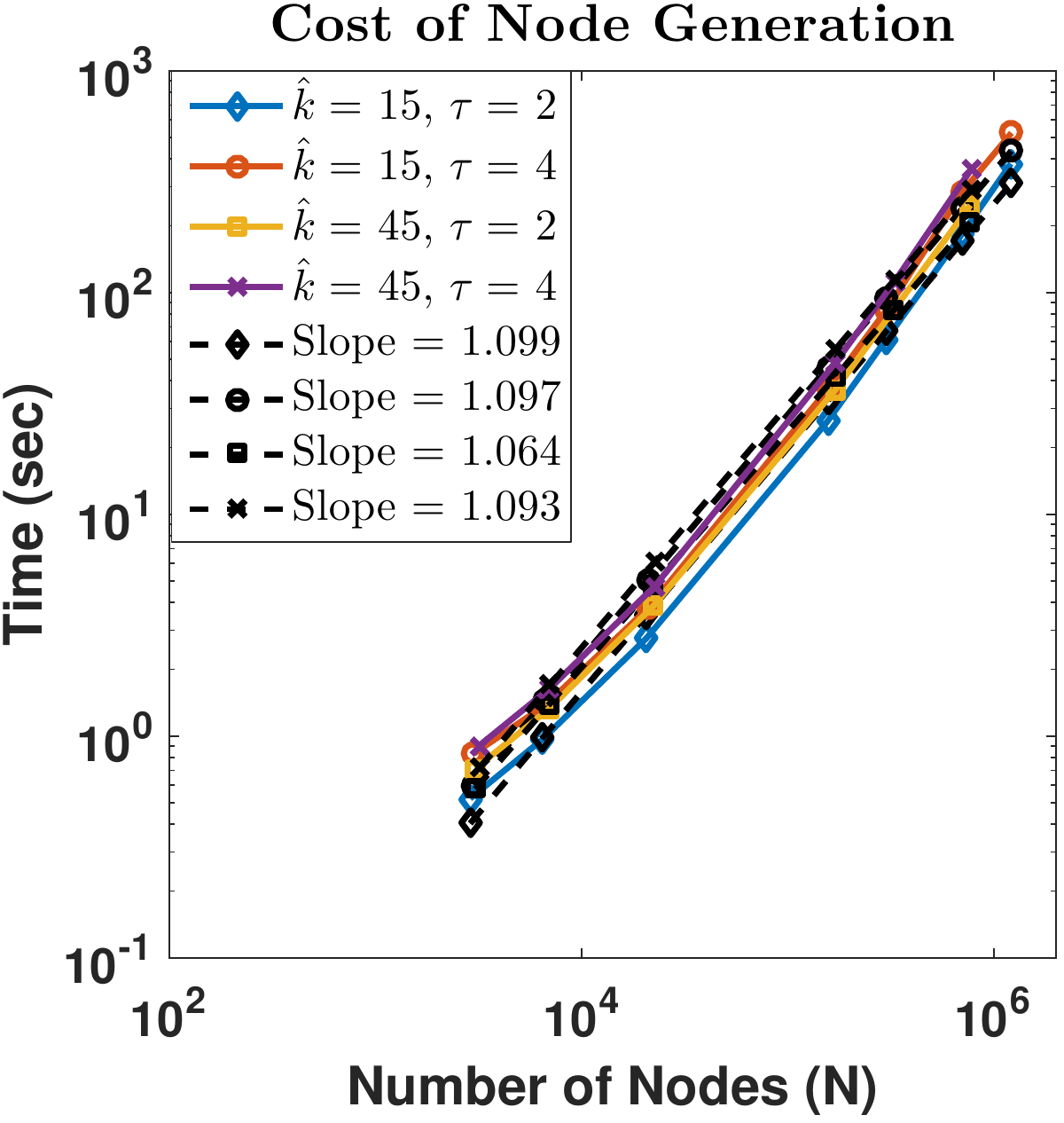} 	
	\label{fig:s1d}
}
\caption{Cost of node generation using Algorithm \ref{alg:node_gen_1} on 2D (top row) and 3D (bottom row) irregular domains (on a loglog plot). The timings are shown as a function of the Poisson neighborhood size $\hat{k}$, the supersampling parameter $\tau$, the number of data sites $N_d$, and the total number of nodes $N$. The dashed lines are lines of best fit used to measure slopes.}
\label{fig:scaling1}	
\end{figure}
\begin{figure}[h!]
\centering
\subfloat[Star with embedded ellipse]
{
	\includegraphics[scale=0.5]{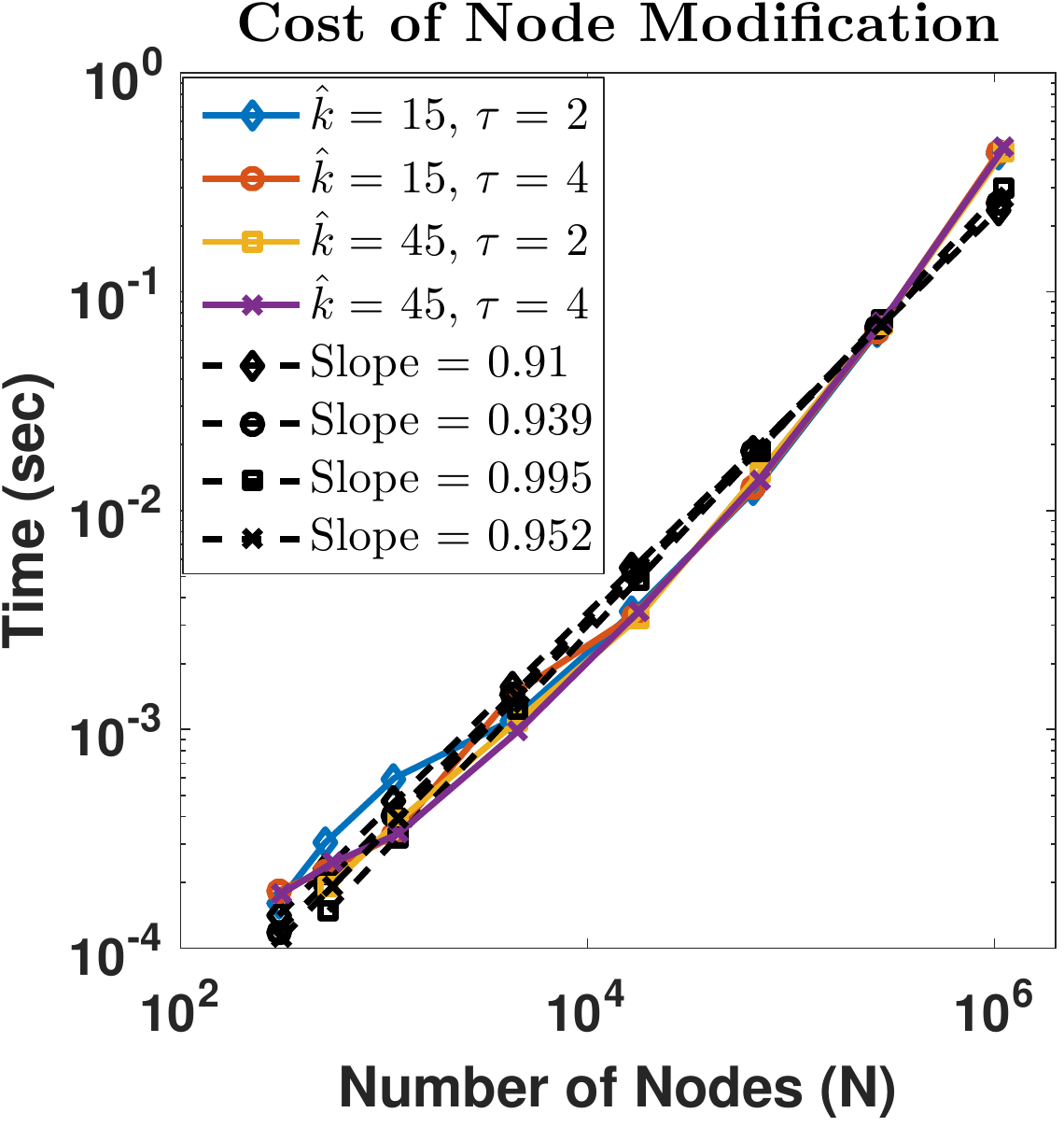}
	
	\label{fig:s2a}
}
\subfloat[Bumpy sphere with embedded red blood cell]
{
	\includegraphics[scale=0.5]{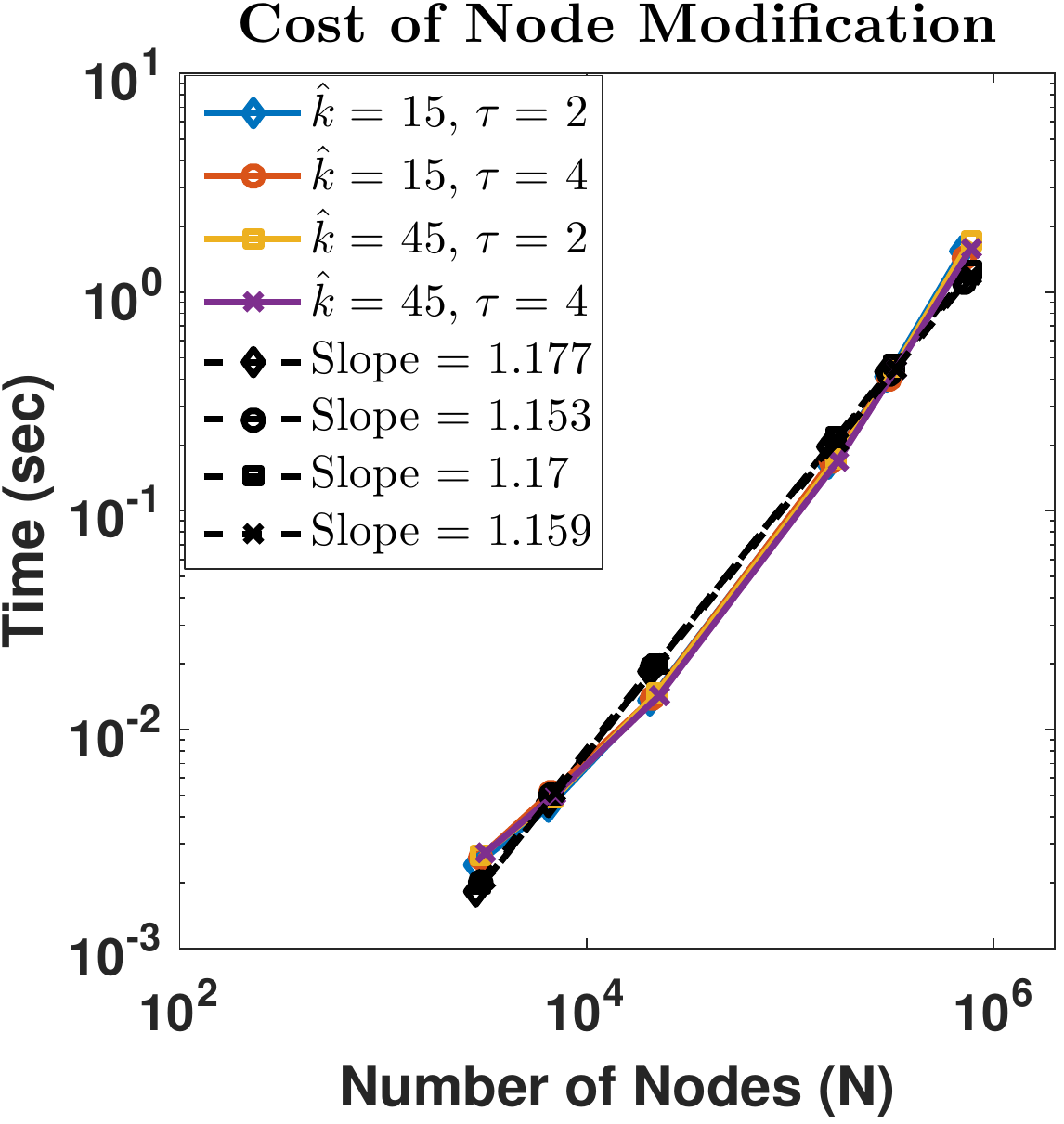} 	
	\label{fig:s2b}
}
\caption{Cost of node set \emph{modification} using Algorithm \ref{alg:node_mod} on 2D (left) and 3D (right) irregular domains (on a loglog plot). The number of inner boundary data nodes is fixed to $N_d^{\Gamma} = 24$ in 2D and $N_d^{\Gamma} = 200$ in 3D. The timings are shown as a function of the Poisson neighborhood size $\hat{k}$, the supersampling parameter $\tau$, and the total number of nodes $N$ (measured prior to modification). The dashed lines are lines of best fit used to measure slopes. }
\label{fig:scaling2}	
\end{figure}

We now present scaling results (in the form of timings) for our node generator. All timings were done using a C/C++ code on the six-core Intel Coffee Lake 8700-K (12 logical cores with hyperthreading) clocked at 4.56 GHz, with 16 GB of 2600 MHz DDR4 RAM. We focus on the star domain from Section 3 and the bumpy sphere domain from~\cite{FuselierWright2013,SWFKJSC2014}. We perform two experiments: the first measures the cost of node generation in these two irregular domains, while the second measures the cost of modifying these node sets with embedded inner boundaries. The results of the first experiment are shown in Figure \ref{fig:scaling1}, and the results from the second experiment are shown in Figure \ref{fig:scaling2}.

Figure \ref{fig:scaling1} verifies that the cost of node generation is indeed approximately $O(N)$ in both 2D (Figure \ref{fig:s1a}) and 3D (Figure \ref{fig:s1b}. The cost in each case goes up slightly as $N_d$, $\hat{k}$, and $\tau$ are increased (though this is hard to see from the graphs), but the slopes remain approximately linear. Code profiling shows that the bulk of the time is taken up by kd-tree operations for large $N$. We also verified that the geometric modeling costs scaled as $O(N_b N_d)$ (not shown).  We return to this in the discussion section.
For the next experiment, we place a single ellipse tilted at an angle of $\pi/4$ with respect to the $x$-axis at the center of the star domain; for the 3D analogue of this experiment, we place a single red blood cell (RBC) in the interior of the bumpy sphere domain. We then measure the costs of modifying the node sets as the node spacing $h$ is decreased and $N$ is increased. It is important to note that this also has the effect of increasing the number of nodes on the boundaries of these inner embedded objects. The results are shown in Figure \ref{fig:scaling2} for different values of $\hat{k}$ and $\tau$. Both the 2D and 3D experiments (Figures \ref{fig:s2a} and \ref{fig:s2b}, respectively) show once again that our asymptotic estimates hold true: we see that the cost of node set modification is approximately $O(N)$ (see figure captions). Obviously, the precise cost depends on the shape, volume, and number of the inner embedded boundaries. Nevertheless, the experiment verifies that one can indeed modify these node sets much more quickly than one can generate them, reflecting the local nature of our node set modification algorithm. From these results, the results from Sections \ref{sec:res_hist1}--\ref{sec:res_err}, we conclude that it is reasonable to set $\tau = 2$ and $\hat{k} = 15$ for our applications. It may be reasonable to adapt $\tau$ according to the local curvature of the domain boundary. We leave this for future work.

%% file: Results.tex
\subsection{Some results}
\label{sec:results}

We now study three aspects of our algorithm. First, we study the spatial distributions of domain boundary nodes obtained from Algorithm \ref{alg:ss_decim}. Next, we explore the influence of the Poisson neighborhood size $\hat{k}$ on the spatial distributions of interior nodes obtained from Algorithm \ref{alg:pds}. Finally, we explore the effect of the Poisson neighborhood size $\hat{k}$ on the stability and errors in an RBF-FD discretizations.

\subsubsection{Spatial distribution of boundary nodes}
\label{sec:res_hist1}
\begin{figure}[h!]
\centering
\subfloat[Naive parametric sampling]
{
	\includegraphics[scale=0.2]{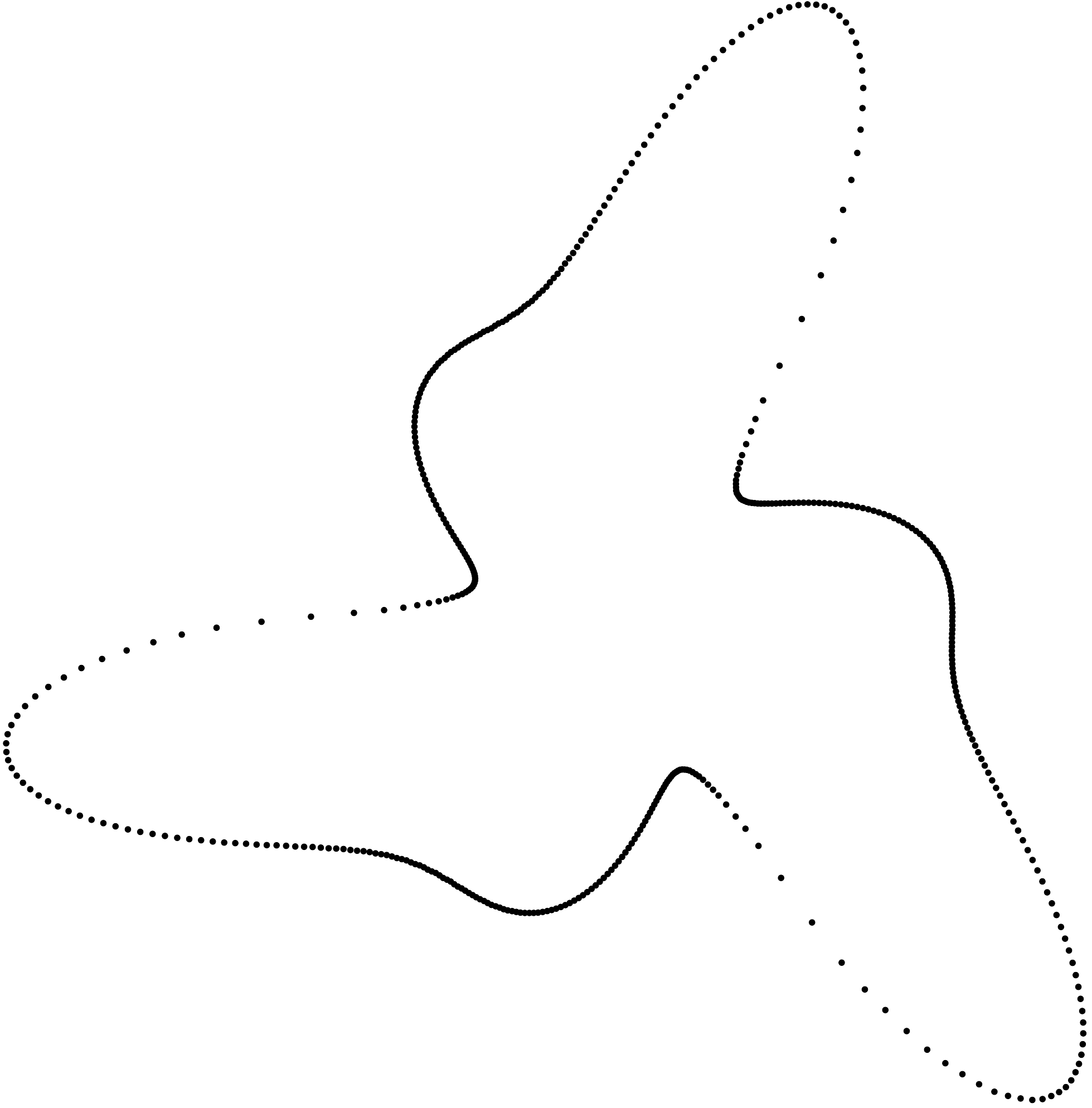} 	
	\label{fig:ps1a}
}
\subfloat[Sampling by Algorithm \ref{alg:ss_decim}]
{
	\includegraphics[scale=0.2]{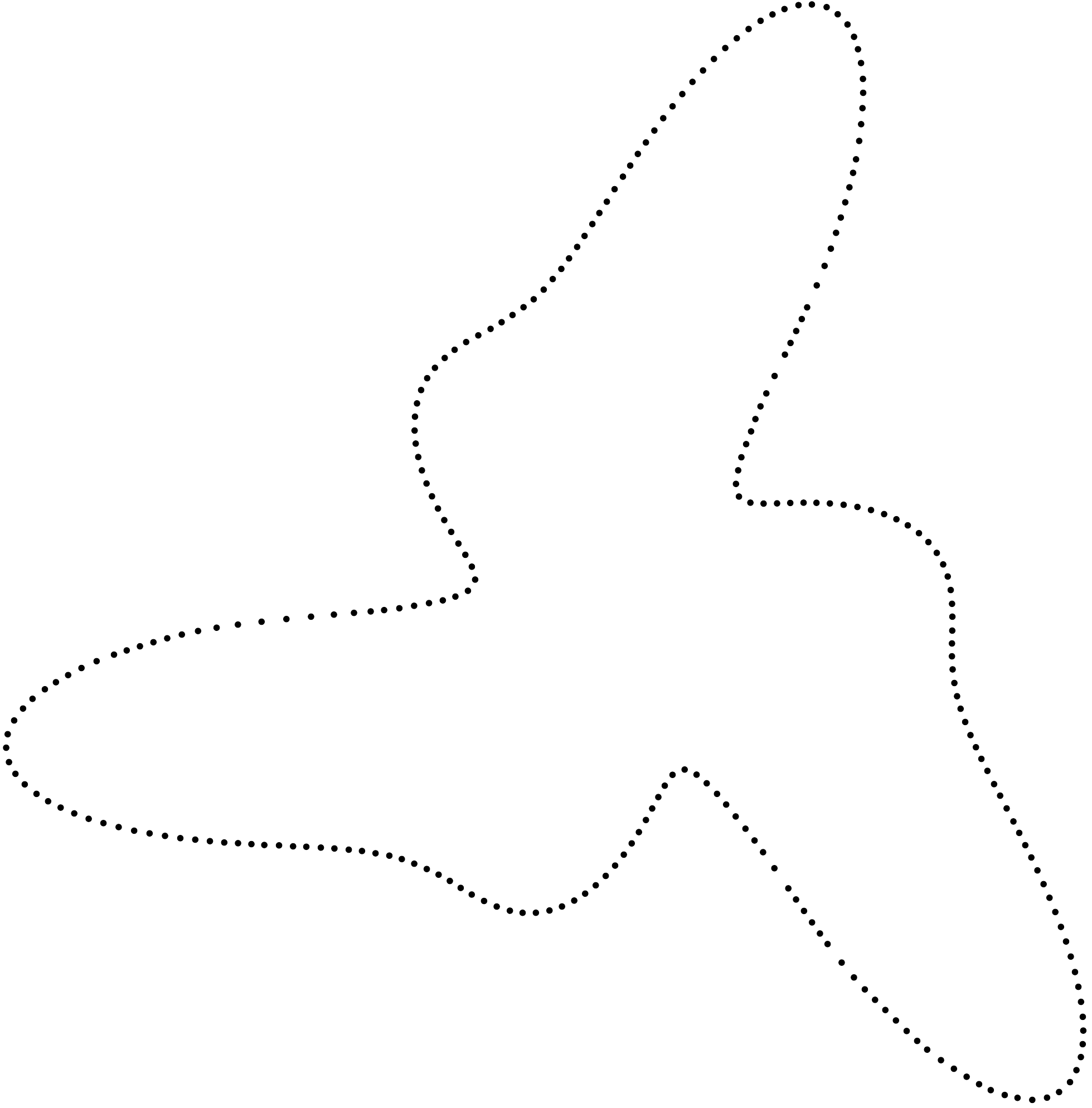}
	
	\label{fig:ps1b}
}

\subfloat[Histogram for $h=0.005$ (naive sampling)]
{
	\includegraphics[scale=0.4]{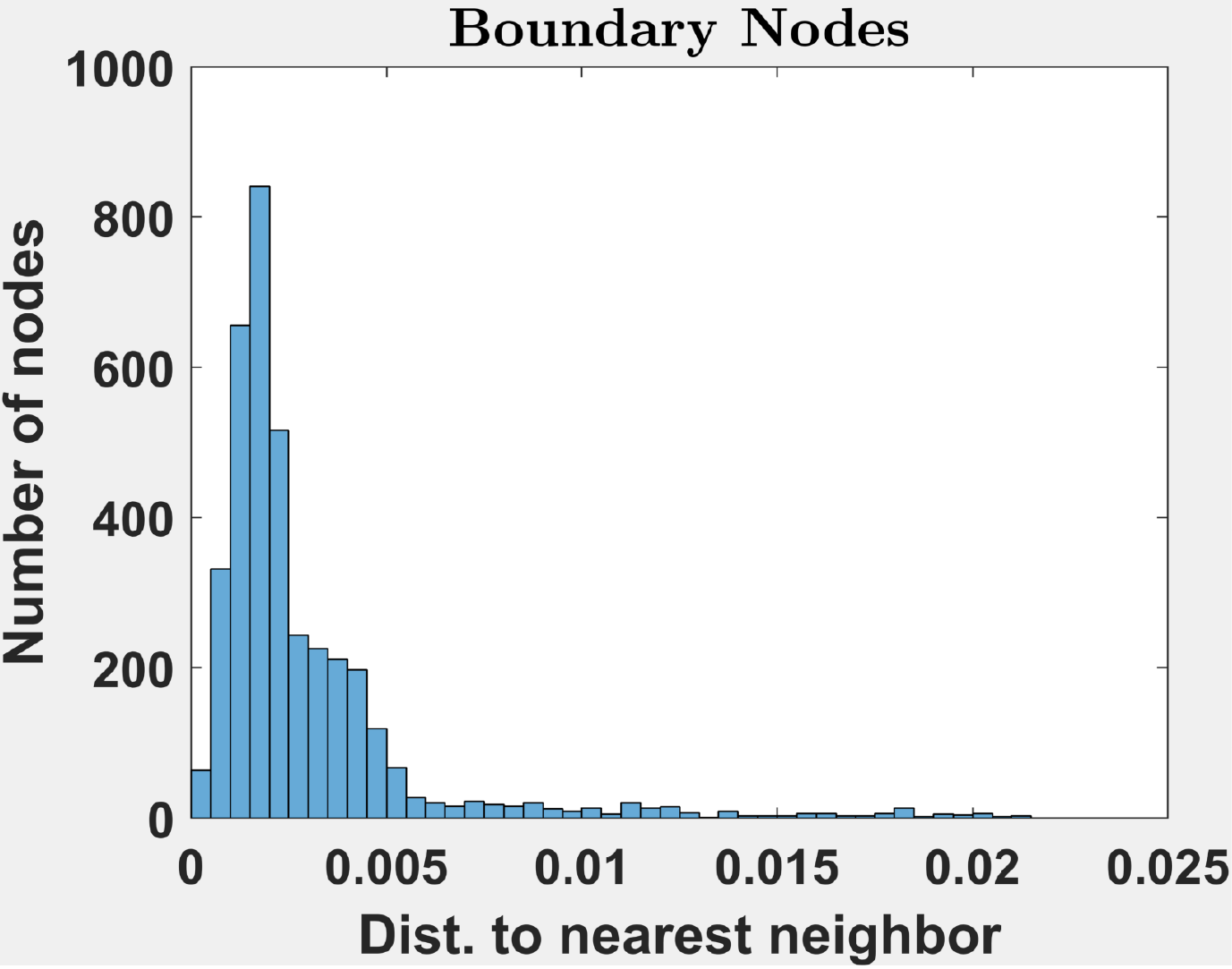} 	
	\label{fig:ps1c}
}
\subfloat[Histogram for $h=0.005$ (Algorithm \ref{alg:ss_decim})]
{
	\includegraphics[scale=0.4]{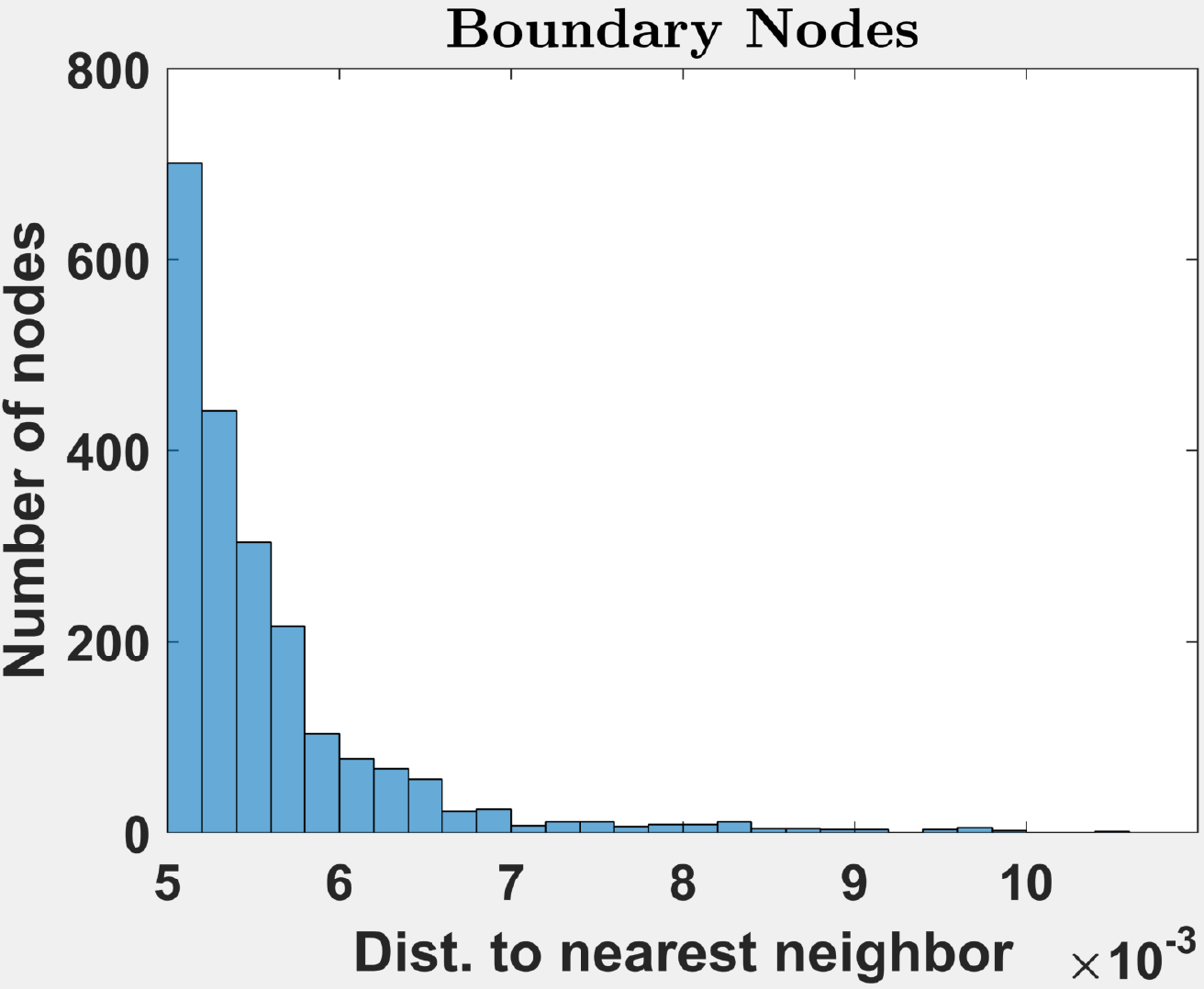}
	
	\label{fig:ps1d}
}
\caption{Comparison of surface sampling techniques for the star domain using supersampling parameter $\tau = 2$.}
\label{fig:poisson_surf1}	
\end{figure}
\begin{figure}[h!]
\centering
\subfloat[Naive parametric sampling]
{
	\includegraphics[scale=0.3]{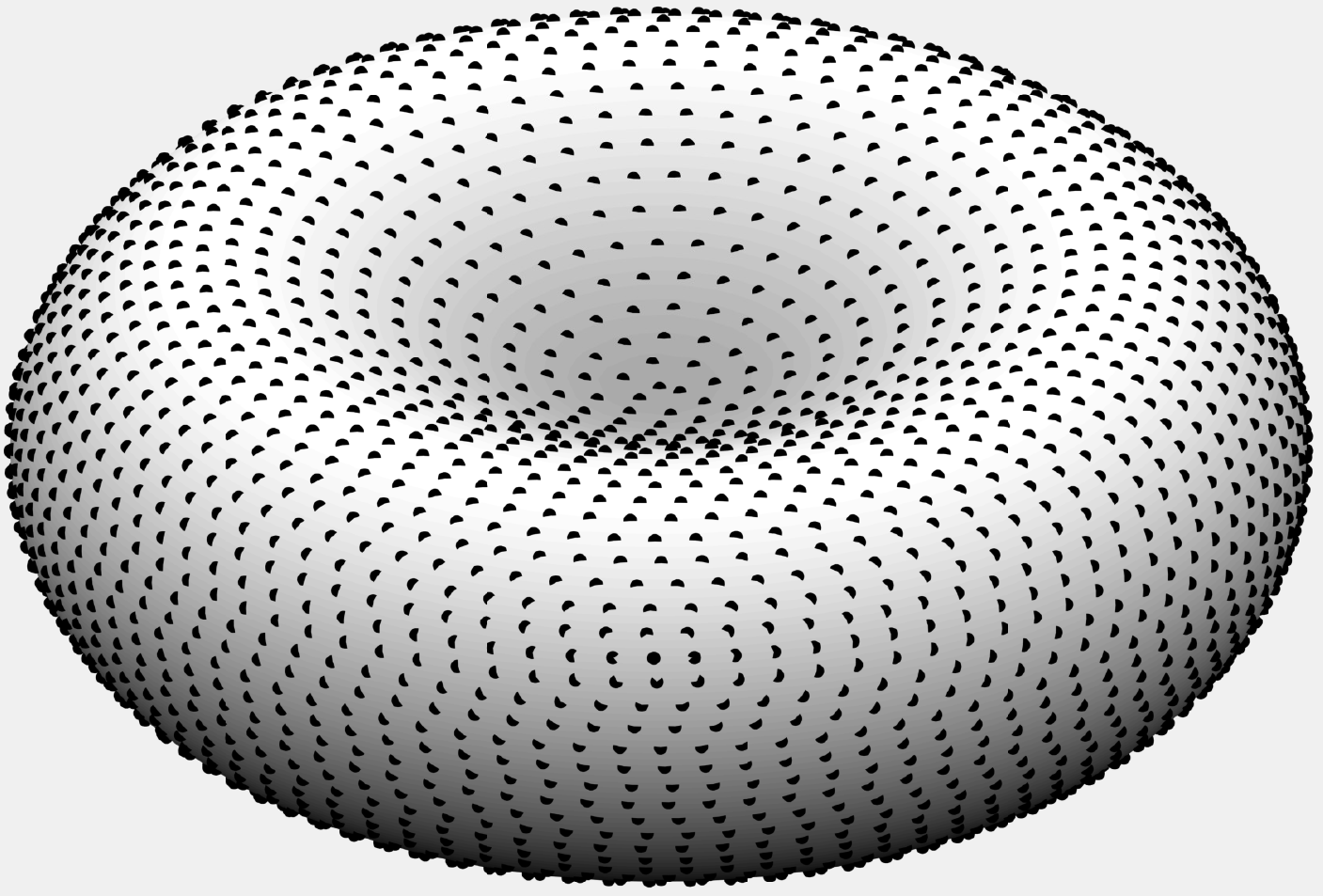} 	
	\label{fig:ps2a}
}
\subfloat[Sampling by Algorithm \ref{alg:ss_decim}]
{
	\includegraphics[scale=0.3]{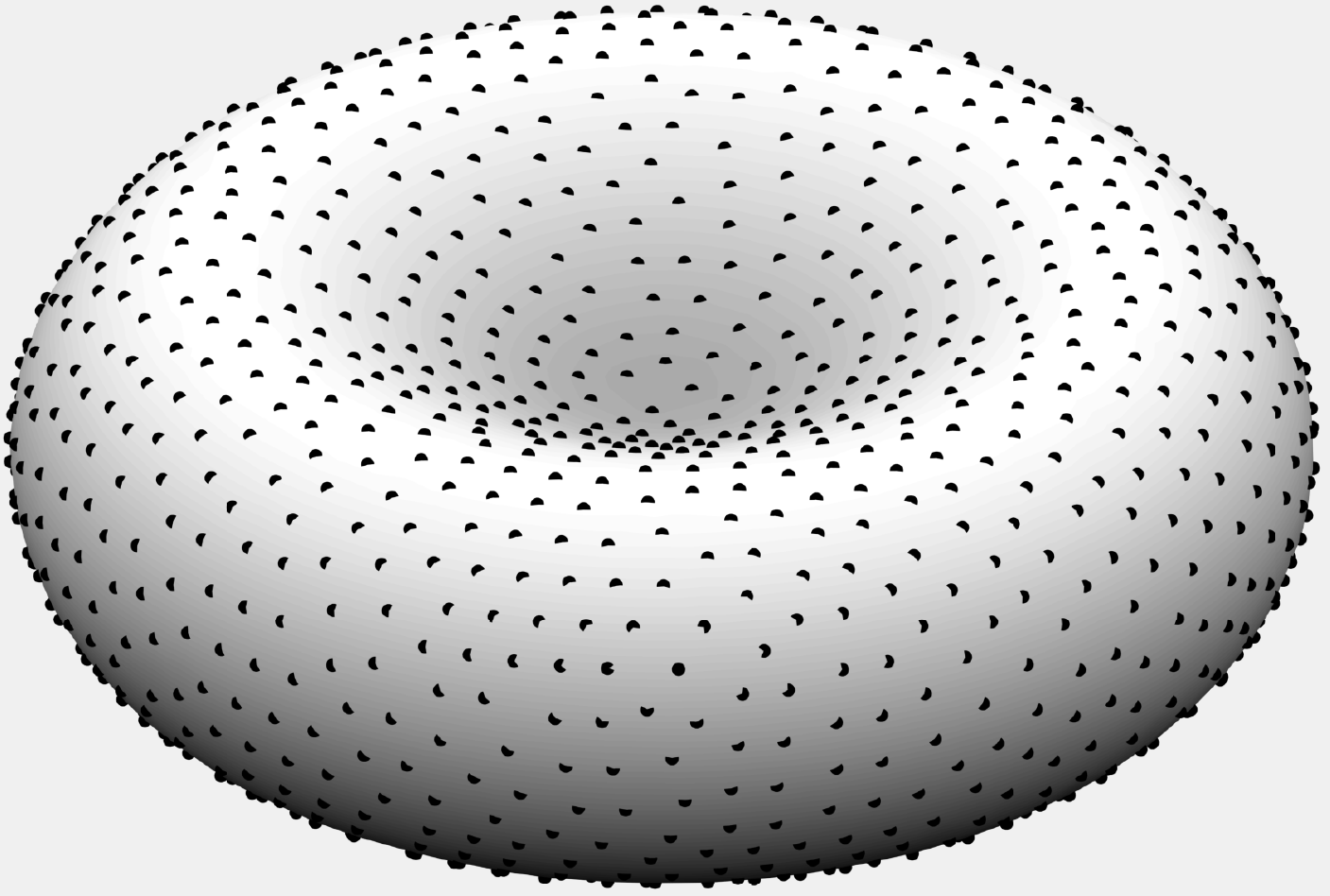} 	
	\label{fig:ps2b}
}

\subfloat[Histogram for $h=0.05$ (naive sampling)]
{
	\includegraphics[scale=0.4]{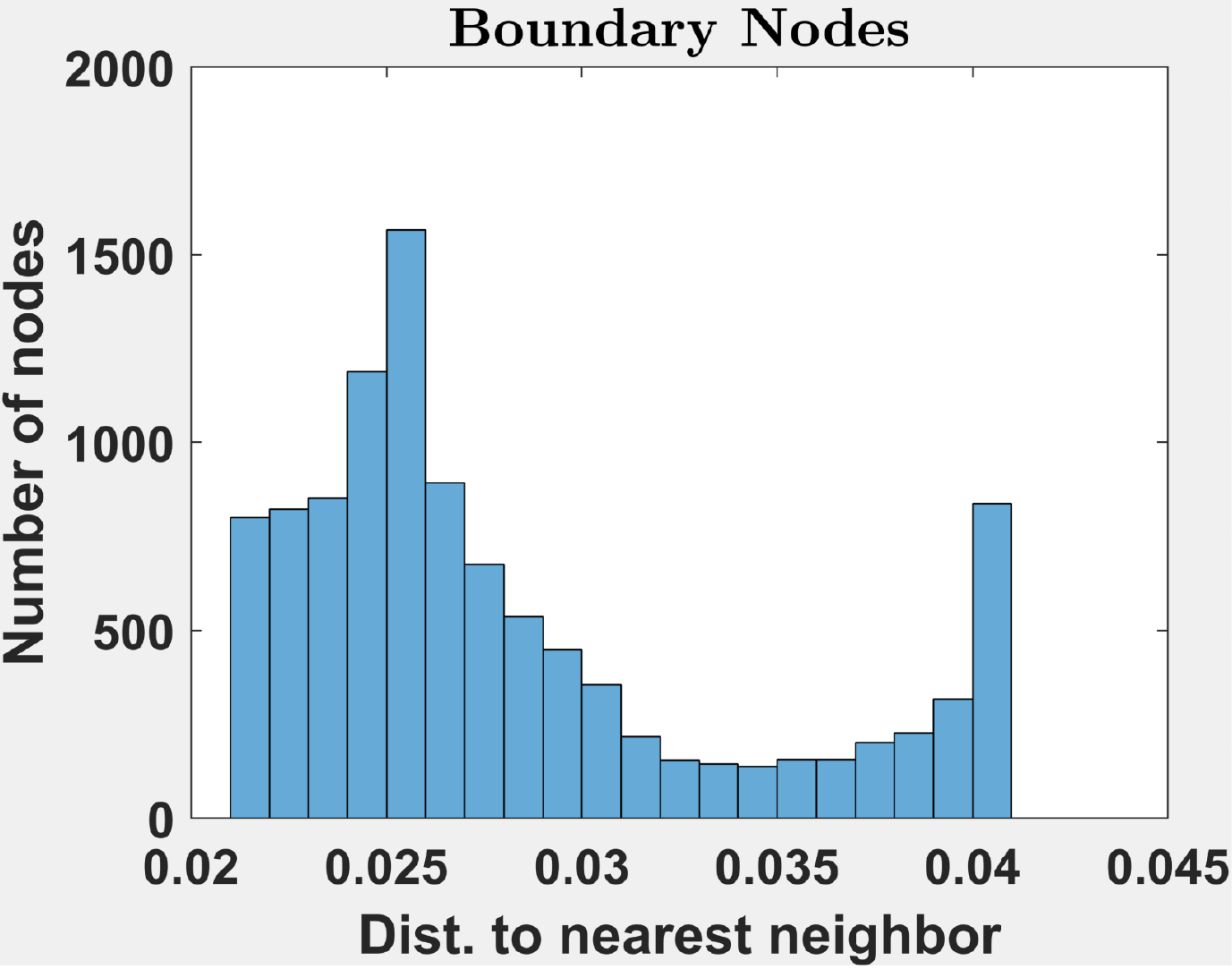} 	
	\label{fig:ps2c}
}
\subfloat[Histogram for $h=0.05$ (Algorithm \ref{alg:ss_decim})]
{
	\includegraphics[scale=0.4]{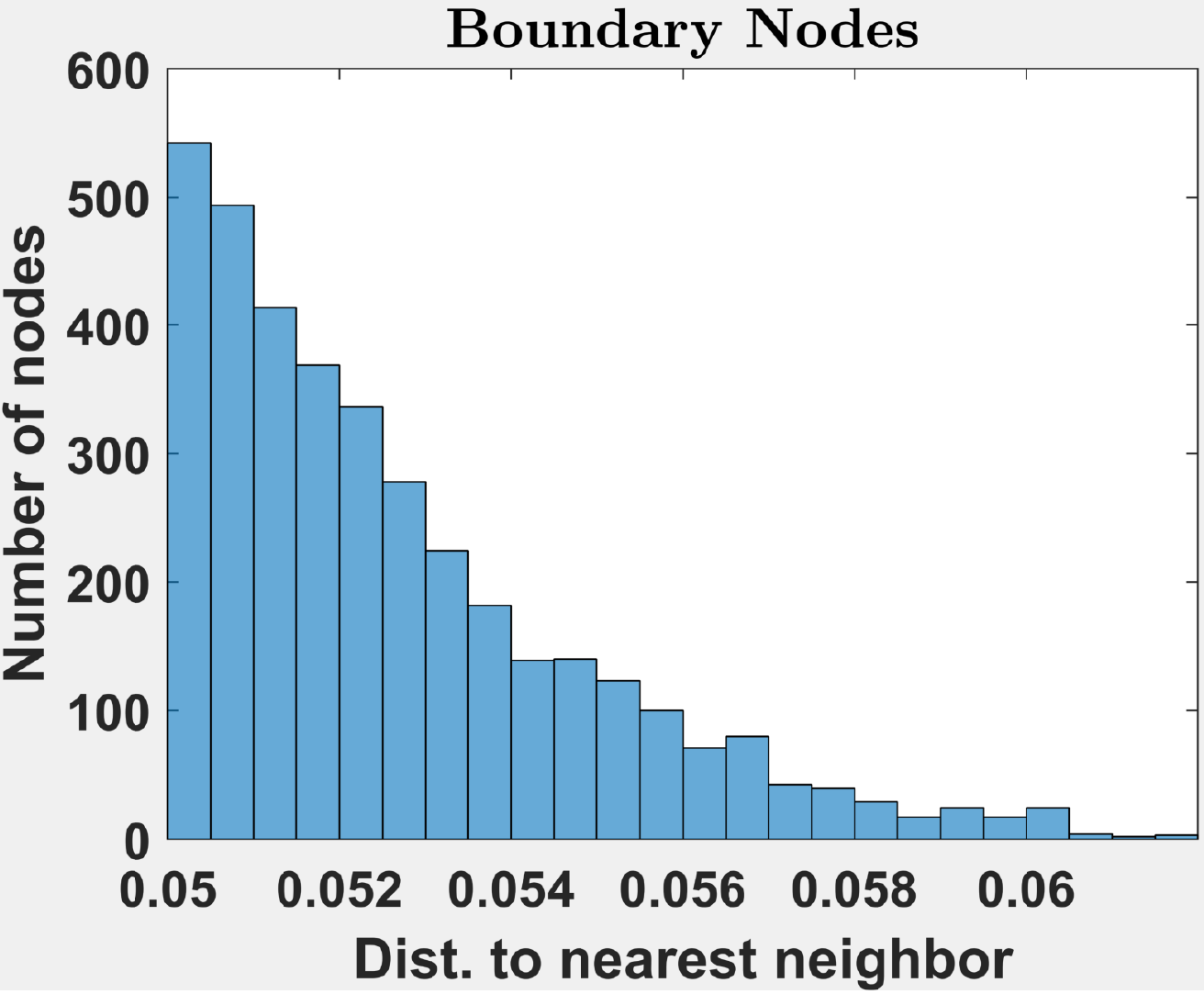}	
	\label{fig:ps2d}
}
\caption{Comparison of surface sampling techniques for the red blood cell domain using supersampling parameter $\tau = 2$.}
\label{fig:poisson_surf2}	
\end{figure}

We first explore the effect of Algorithm \ref{alg:ss_decim} on the node distributions produced on surfaces, and compare it to a naive sampling of the parametric map. To briefly illustrate the efficacy of this procedure, we sample two functions (corresponding to domain boundary shapes) whose maps from parametric space to Cartesian space are likely to produce distorted point sets. The 2D shape (the star domain) is given by the following $C^0$ function:
\begin{align}
r(\lambda) = |\cos(1.5 \lambda)|^{\sin(3 \lambda)}, \\
x(\lambda) = r(\lambda)\cos(\lambda), y(\lambda) = r(\lambda)\sin(\lambda),
\end{align} 
where $\lambda \in [0,2\pi)$. The 3D shape is a classic Red Blood Cell (RBC) shape, obtained by smoothly distorting the sphere, and is therefore $C^{\infty}$~\cite{FuselierWright2013,LSWSISC2017}. For this test, we interpolate the star shape with the SBF interpolant at $N_d = 128$ points (due to its low smoothness) and attempt to generate Cartesian samples with an average node spacing of $h = 0.005$. We interpolate the 3D shape with the SBF interpolant at $N_d = 700$ points, and attempt to generate Cartesian samples with an average node spacing of $h = 0.05$. The resulting node sets and their histograms are shown in Figures \ref{fig:poisson_surf1} and \ref{fig:poisson_surf2}. 
Figure \ref{fig:ps1a} clearly shows the clustering in the naive sampling technique for the star domain, corroborated by the histogram in Figure \ref{fig:ps1c} which shows a large number of nodes with a distance of $h=0.0025$ to the nearest neighbor (rather than the desired $h=0.005$). Examining the histogram, we also see that the nodes are also sometimes very far apart (as far apart as $h=0.02$). In contrast, Algorithm \ref{alg:ss_decim} produces much more uniform node sets, which can be verified visually (Figure \ref{fig:ps1b}) and from the corresponding histogram (Figure \ref{fig:ps1d}). In this case, the histogram shows the most number of nodes in the bin corresponding to $h=0.005$, with exponentially fewer nodes in bins corresponding to larger $h$ values. The 3D results are shown in Figures \ref{fig:ps2a} and \ref{fig:ps2b}. Figure \ref{fig:ps2c} (naive sampling) shows that most nodes are in a bin close to $h=0.03$, but are distributed over a relatively wide range. With naive sampling, there appear to be no nodes whatsoever in the desired $h=0.05$ bin! In marked contrast, the histogram in Figure \ref{fig:ps2d} (corresponding to sampling by Algorithm \ref{alg:ss_decim}) shows the most number of nodes in the $h=0.05$ bin (as desired), with exponentially fewer nodes in bins corresponding to larger $h$ values. In all cases, Algorithm \ref{alg:ss_decim} still appears to produce quasi-uniform node sets.

\subsubsection{Spatial distribution of interior nodes}
\label{sec:res_hist2}
\begin{figure}[h!]
\centering
\subfloat[Star, $\hat{k}=15$, $h=0.005$]
{
	\includegraphics[scale=0.4]{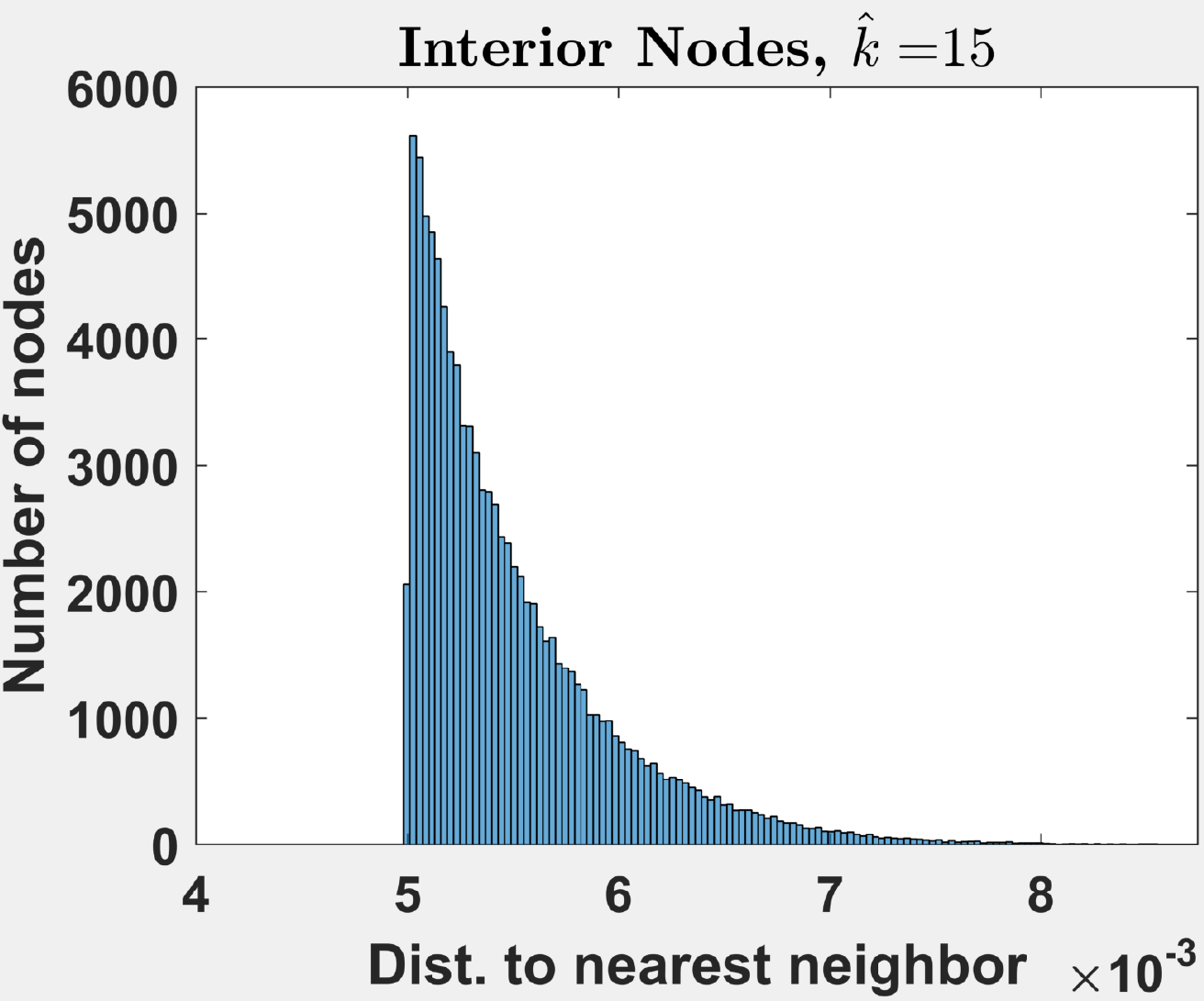}
	
	\label{fig:h1a}
}
\subfloat[Star, $\hat{k}=45$, $h=0.005$]
{
	\includegraphics[scale=0.4]{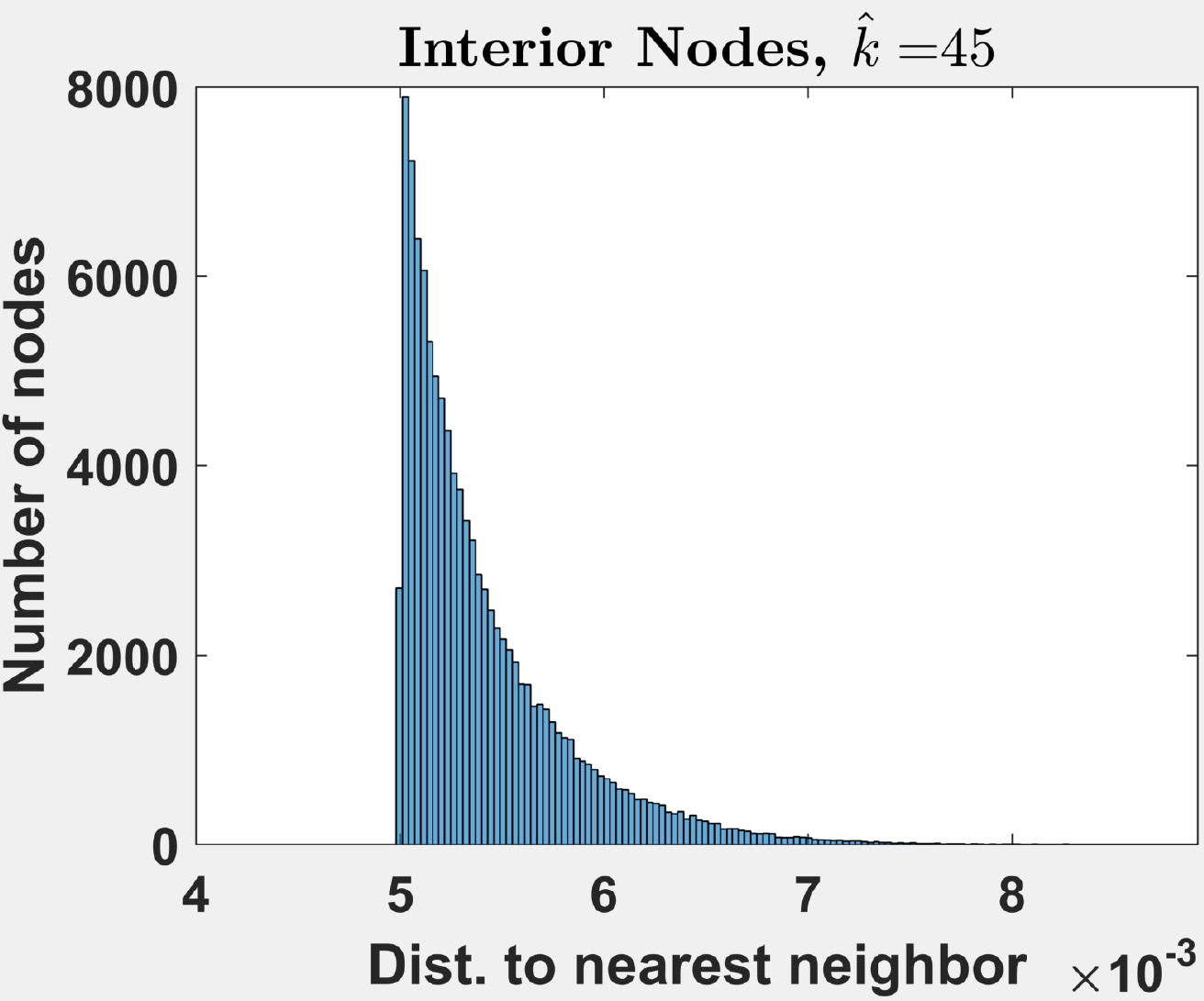} 	
	\label{fig:h1b}
}

\subfloat[Red blood cell, $\hat{k}=15$, $h=0.05$]
{
	\includegraphics[scale=0.4]{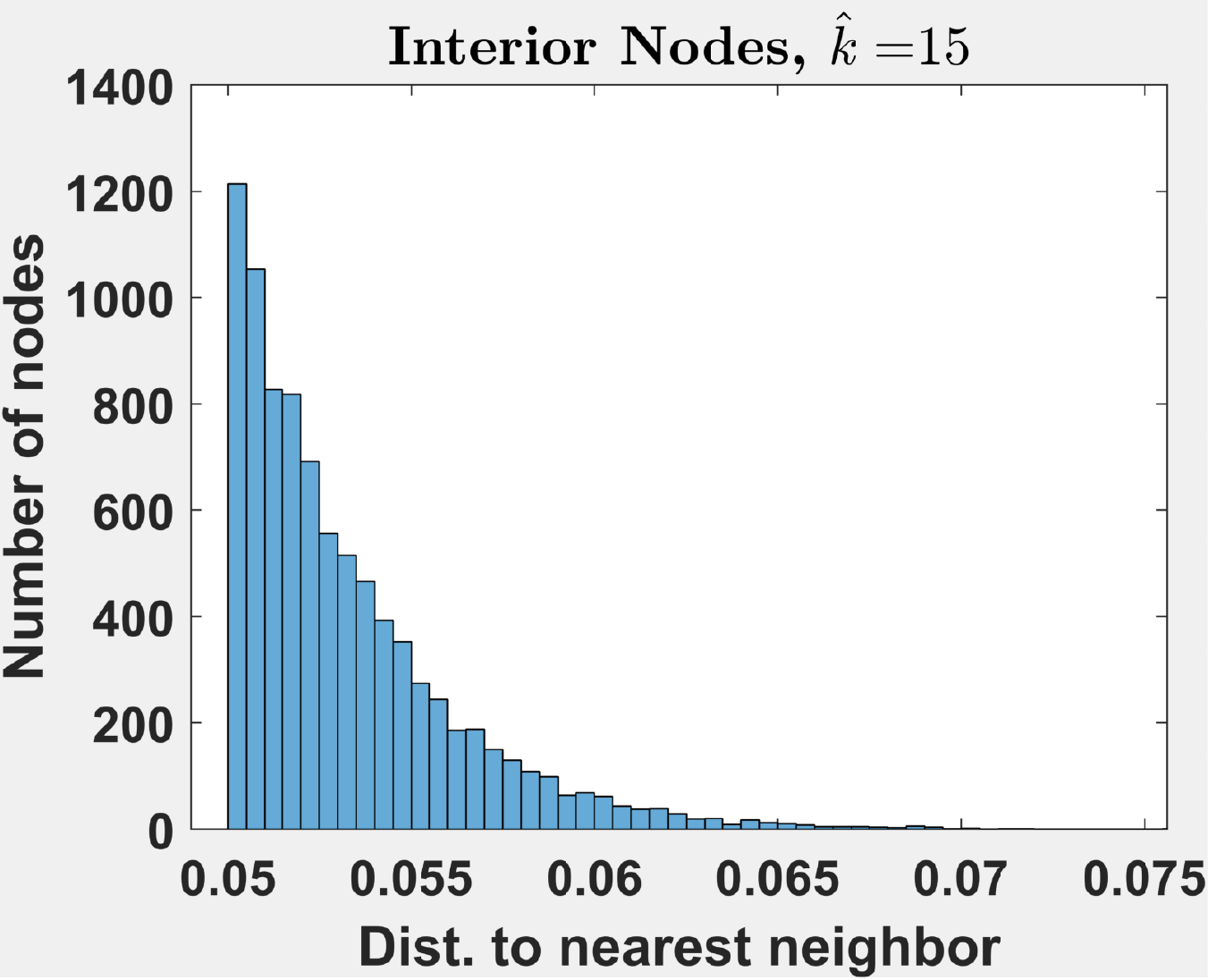}
	
	\label{fig:h1c}
}
\subfloat[Red blood cell, $\hat{k}=45$, $h=0.05$]
{
	\includegraphics[scale=0.4]{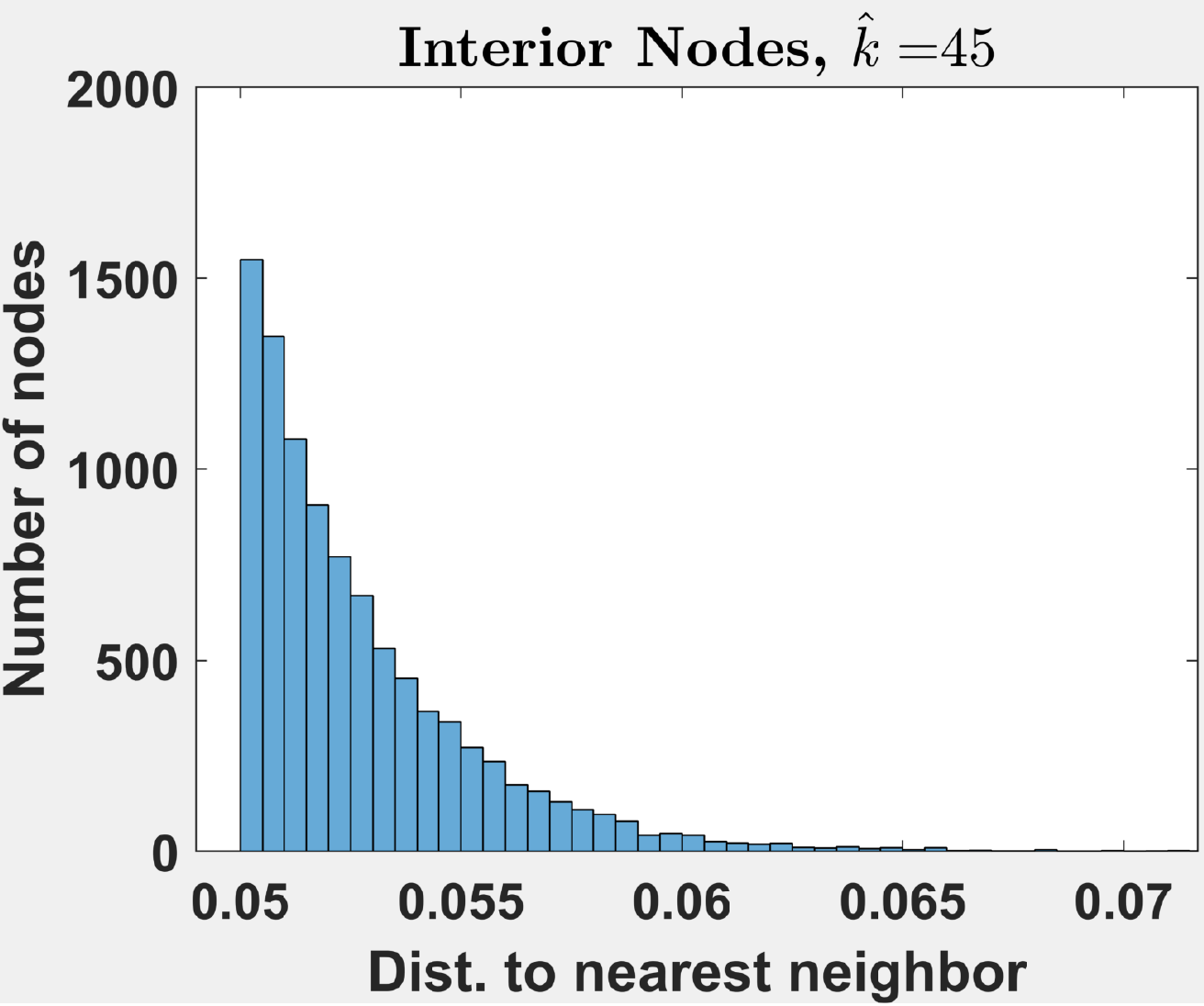} 	
	\label{fig:h1d}
}
\caption{Histograms of node sets generated by Algorithm \ref{alg:node_gen_1} in the interiors of the star domain (top row) and the red blood cell domain (bottom row) as a function of the Poisson neighborhood size $\hat{k}$.}
\label{fig:hist2d1}	
\end{figure}
We turn our attention to the node sets produced by Algorithm \ref{alg:node_gen_1} in the interiors of domains. Recall that this is accomplished by applying Algorithm \ref{alg:pds} in the domain bounding box to obtain $N_{obb}$ nodes, then eliminating those nodes that lie outside the parametric SBF interpolant to the boundary. It is important to note that Algorithm \ref{alg:pds} utilizes the Poisson neighborhood size $\hat{k}$ to control the cost of sampling. However, $\hat{k}$ also determines the number of nodes against which samples are compared to specify the desired separation distance $h$. Indeed, one could imagine setting $\hat{k} = N_{obb}$ to obtain a perfectly-spaced set of nodes at a cost of $O(N_{obb}^2)$ for node generation. Since $\hat{k}$ controls both computational cost and spatial distribution, it is important to explore the effect of this parameter on spatial distributions of nodes. To do so, we compute histograms of distances to the nearest neighbor for node sets in the interior of the star domain with $h=0.005$ (2D), and the red blood cell with $h=0.05$ (3D). The results are shown in Figure \ref{fig:hist2d1}. From Figure \ref{fig:hist2d1}, it is clear that Algorithm \ref{alg:node_gen_1} produces node sets that are quasi-uniform regardless of the $\hat{k}$ value used. Most nodes lie in the desired bins of $h=0.005$ (in 2D) and $h=0.05$ (in 3D), and the number of nodes in bins corresponding to larger values of $h$ drops off exponentially in both 2D and 3D. Very similar node distributions were obtained when objects were embedded within (results not shown).

\subsubsection{Stability for RBF-FD discretizations}
\label{sec:res_stab}
\begin{figure}[h!]
\centering
\subfloat[Eigenvalues on Gmsh nodes]
{
	\includegraphics[scale=0.45]{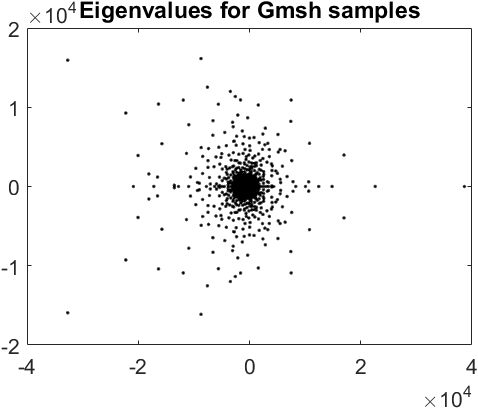}
	
	\label{fig:nsa}
}
\subfloat[Eigenvalues on Algorithm \ref{alg:node_gen_1} nodes]
{
	\includegraphics[scale=0.45]{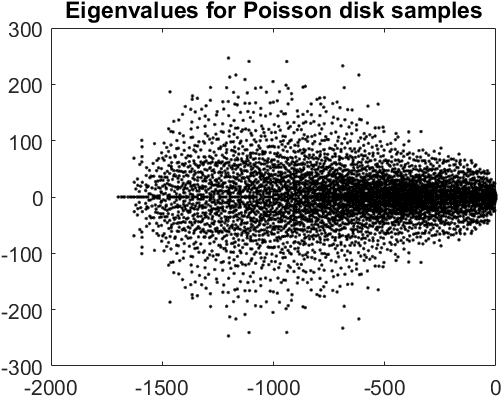} 	
	\label{fig:nsb}
}
\caption{Eigenvalues of the discrete Laplacian formed by a sixth-order RBF-FD method in the unit ball for nodes obtained from Gmsh (left) and Algorithm \ref{alg:node_gen_1} with $\hat{k}=15$ and $\tau=2$ (right).}
\label{fig:node_stab}	
\end{figure}
Our motivation for designing Algorithm \ref{alg:node_gen_1} was at least partly because nodes obtained from popular node generators such as Gmsh were not always suitable for RBF-FD discretizations. Figure \ref{fig:node_stab} shows an example of this. A sixth-order RBF-FD discretization of the Laplacian in the unit ball resulted in eigenvalues with positive real parts when using a Gmsh-generated node set with $N=4561$ nodes (Figure \ref{fig:nsa}). In contrast, the same high-order discretization on $N=5157$ nodes generated by Algorithm \ref{alg:node_gen_1} results in a well-behaved spectrum (Figure \ref{fig:nsb}) with a relatively small spread even along the imaginary axis. We note that it may indeed be possible to obtain appropriate nodes for RBF-FD discretizations from Gmsh if the right parameters and algorithms are used. However, Algorithm \ref{alg:node_gen_1} required no fine tuning for this example.

\subsubsection{Errors in RBF-FD discretizations}
\label{sec:res_err}
\begin{figure}[h!]
\centering
\subfloat[Convergence rates for $\hat{k}=15$]
{
	\includegraphics[scale=0.5]{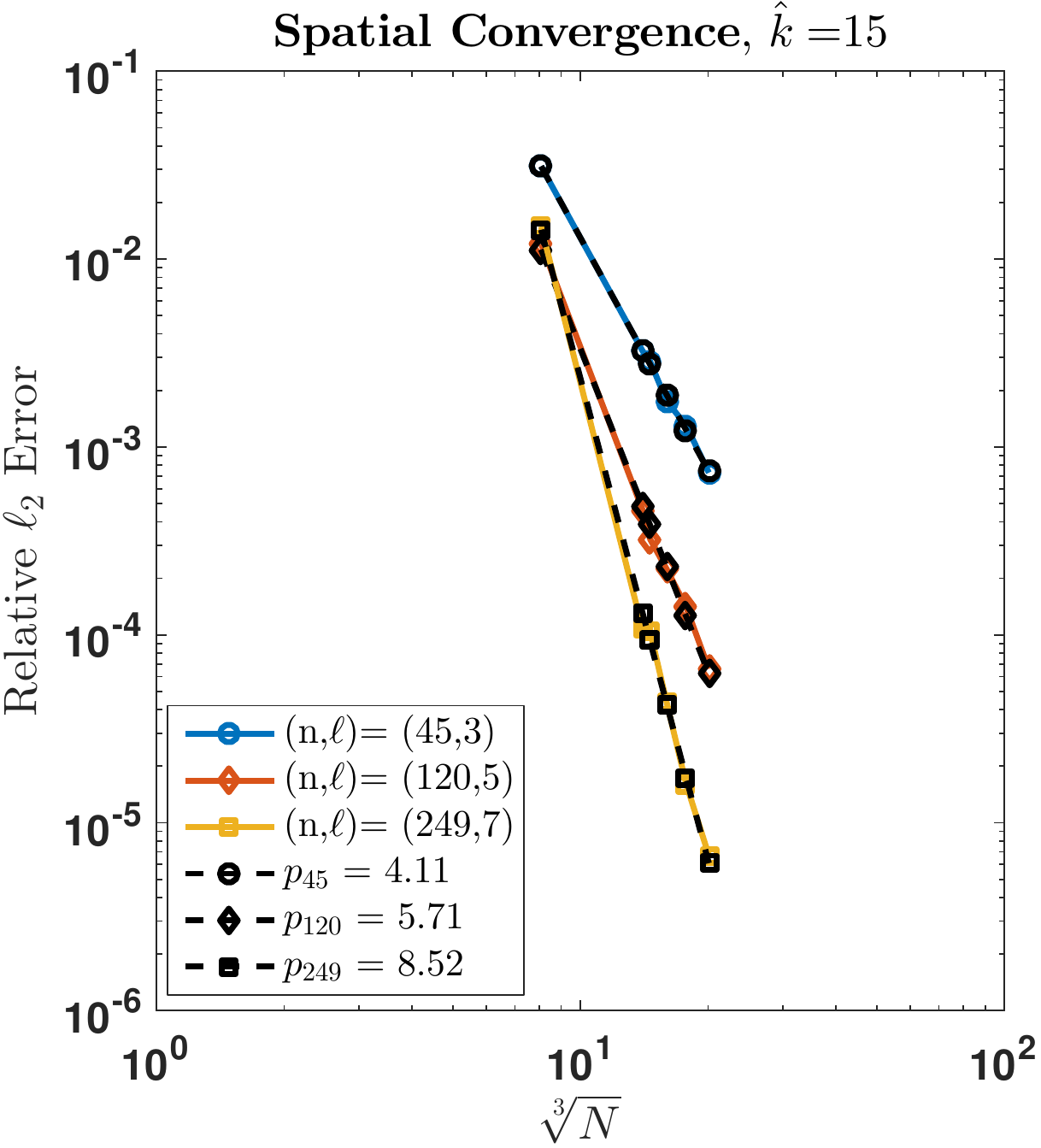}
	
	\label{fig:csa}
}
\subfloat[Convergence rates for $\hat{k}=45$]
{
	\includegraphics[scale=0.5]{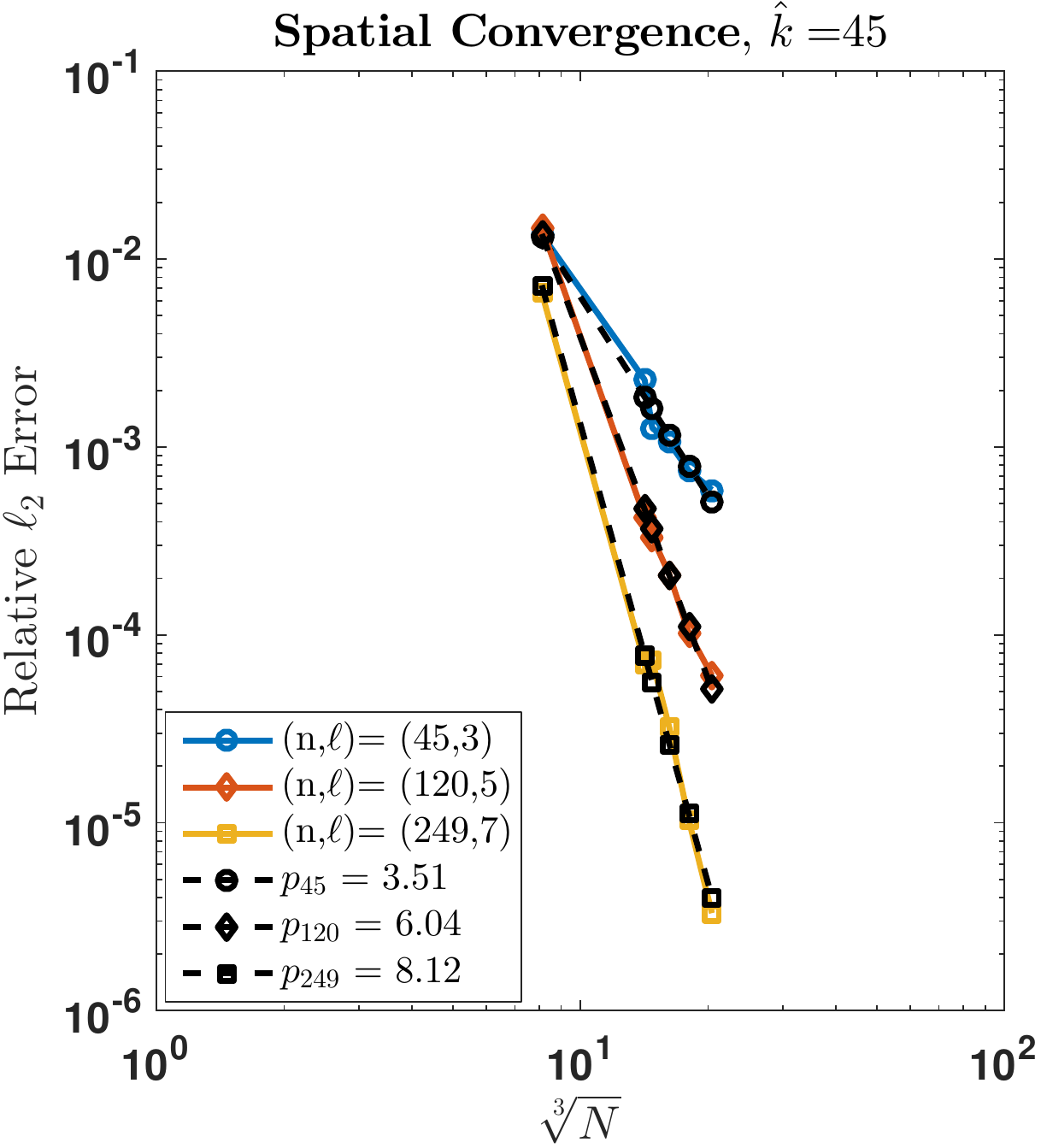} 	
	\label{fig:csb}
}
\caption{Errors in the solution to \eqref{eq:heat1}--\eqref{eq:heat2} in the red blood cell domain ($N_d = 800$) as a function of the stencil size $n$, number of nodes $N$, polynomial degree $\ell$, and Poisson neighborhood size $\hat{k}$.}
\label{fig:conv_study}	
\end{figure}
The goal of this article was to design a node generator that produces node sets suitable for RBF-FD discretizations. In this section, we provide some evidence as to that suitability by exploring the effects of the parameter $\hat{k}$ on the errors and convergence rates from RBF-FD discretizations of the heat equation:
\begin{align}
\frac{\partial c(\vx,t)}{\partial t} &= \Delta c + f(\vx,t), \vx \in \Omega,\label{eq:heat1}\\
\frac{\partial c(\vx,t)}{\partial \vn} &= g(\vx,t), \vx \in \gamma, \label{eq:heat2}
\end{align}
where $\Omega$ is the red blood cell domain seen earlier, and $\gamma$ is its curved boundary. We have restricted ourselves to a 3D test for brevity. We use $N_d = 800$ seed nodes as the starting point for the node generator; this is sufficient to ensure that the errors in geometric modeling match the high orders of accuracy in RBF-FD. To measure convergence rates, we manufacture a solution to the heat equation by prescribing $c(\vx,t)$, then computing a corresponding $f(\vx,t)$ and $g(\vx,t)$ that make the prescribed solution hold. We use the infinitely-smooth ansatz:
\begin{align}
c(\vx,t) = c(x,y,z,t) = 1 + \sin(\pi x)\cos(\pi y)\sin(\pi z) e^{-\pi t}.
\end{align}
For the spatial discretization, we use the overlapped RBF-FD method~\cite{ShankarJCP2017}, and the ghost node technique outlined in~\cite{SFJCP2018}. The RBF-FD discretization is achieved by using PHS RBFs in conjunction with high-degree polynomials, where the convergence rates are dictated purely by the polynomial degree~\cite{FlyerPHS,FlyerNS,BarnettPHS}. Since these discretization techniques are explained in detail in the references, we restrict ourselves to a description of our results. Primarily, when increasing the RBF-FD stencil size $n$ and consequently the polynomial degree $\ell$, we expect to see higher orders of convergence. We run these convergence studies and calculate $\ell_2$ errors in the numerical solution for different values of $n$, $\ell$, $\hat{k}$, and $h \propto \frac{1}{\sqrt[3]{N}}$. The results are shown in Figure \ref{fig:conv_study}. Studying Figures \ref{fig:csa} and \ref{fig:csb} closely, we see slightly different convergence rates between the two figures for the same stencil size and polynomial degree. From the data (not shown), the errors for $\hat{k} = 45$ appear to be slightly lower than the errors for $\hat{k} = 15$, but not significantly so at higher values of $N$. This is not surprising, given the similarity in the histograms for these two $\hat{k}$ values, but is reassuring since it indicates that we can use $\hat{k}=15$ and still achieve low errors and high orders of convergence on fairly irregular domains.

%% file: Discussion.tex
\section{Discussion}
\label{sec:summary}

In this article, we presented new algorithms for \emph{one-shot} node generation on irregular domains. The algorithm utilizes a parameter-free high-order boundary representation based on SBF interpolation, with a simple strategy for obtaining quasi-uniform node sets on domain boundaries. When used in conjunction with a bounding box based on PCA, this allowed for the use of Poisson disk sampling without hand-tuning. The simplicity of the algorithm allowed for a straightforward complexity analysis of our algorithms in 2D and 3D. We demonstrated that our algorithms achieve a scaling of $O(N)$ both in theory and in practice. Further, we demonstrated that our algorithm for local node set modification also exhibited a complexity of $O(N)$, albeit with a much smaller constant than in the node generation case. We demonstrated that the node sets were quasi-uniform, and that they were suitable for RBF-FD discretizations even for the value of $\hat{k}=15$ and $\tau=2$.

Profiling revealed that our kd-tree implementation was the primary bottleneck in our algorithm, with the nearest neighbor search and ball query operation costs dominating the cost of node generation. A faster data structure for these operations would likely significantly improve the wall-clock times for node generation and modification. Similarly, a bounding box hierarchy would enable rapid node set modification when hundreds of time-varying embedded boundaries are involved. These are areas of future research. The second significant bottleneck is the high-order boundary representation, with asymptotic costs of $O(N^{1.5})$ in 2D and $O(N^2)$ in 3D. Though the constants in front of these terms are small, it may still be desirable to use a high-order boundary representation that costs $O(N_d)$ operations to obtain; this should also speed up evaluation of the interpolant significantly. The first author plans to pursue this area of research. Finally, the authors plan to leverage this node generation framework to investigate biological problems involving coupled bulk-surface chemical transport.